\documentclass[iop]{emulateapj}
\usepackage[usenames,dvipsnames]{xcolor}
\usepackage{color}
\usepackage{multirow}
\usepackage{longtable}

\shorttitle{Structural evolution in massive galaxies}
\shortauthors{Tadaki et al.}

\begin{document}

\title{Structural evolution in massive galaxies at $z\sim2$}

\author{Ken-ichi Tadaki\altaffilmark{1},
Sirio Belli\altaffilmark{2}, 
Andreas Burkert\altaffilmark{3,4}, 
Avishai Dekel\altaffilmark{5}, 
Natascha M. F\"{o}rster Schreiber\altaffilmark{4}, 
Reinhard Genzel\altaffilmark{4}, 
Masao Hayashi\altaffilmark{1}, 
Rodrigo Herrera-Camus\altaffilmark{6}, 
Tadayuki Kodama\altaffilmark{7}, 
Kotaro Kohno\altaffilmark{8,9}, 
Yusei Koyama\altaffilmark{1}, 
Minju M. Lee\altaffilmark{4}, 
Dieter Lutz\altaffilmark{4}, 
Lamiya Mowla\altaffilmark{10}, 
Erica J. Nelson\altaffilmark{2},
Alvio Renzini\altaffilmark{11}, 
Tomoko L. Suzuki\altaffilmark{7}, 
Linda J. Tacconi\altaffilmark{4}, 
Hannah \"{U}bler\altaffilmark{4}, 
Emily Wisnioski\altaffilmark{12,13},
Stijn Wuyts\altaffilmark{14}
}

\affil{\altaffilmark{1} National Astronomical Observatory of Japan, 2-21-1 Osawa, Mitaka, Tokyo 181-8588, Japan; tadaki.ken@nao.ac.jp}
\affil{\altaffilmark{2} Harvard-Smithsonian Center for Astrophysics, 60 Garden Street, Cambridge, MA 02138, USA}
\affil{\altaffilmark{3} Universit\"{a}ts-Sternwarte Ludwig-Maximilians-Universit\"{a}t (USM), Scheinerstr. 1, M\"{u}nchen, D-81679, Germany}
\affil{\altaffilmark{4} Max-Planck-Insitut f\"{u}r extraterrestrische Physik, Giessenbachstrasse, D-85748 Garching, Germany}
\affil{\altaffilmark{5} Racah Institute of Physics, The Hebrew University, Jerusalem 91904 Israel}
\affil{\altaffilmark{6} Astronomy Department, Universidad de Concepci\'{o}n, Barrio Universitario, Concepci\'{o}n, Chile}
\affil{\altaffilmark{7} Astronomical Institute, Tohoku University, 6-3, Aramaki, Aoba, Sendai, Miyagi, 980-8578, Japan}
\affil{\altaffilmark{8} Institute of Astronomy, School of Science, The University of Tokyo, 2-21-1 Osawa, Mitaka, Tokyo 181-0015, Japan}
\affil{\altaffilmark{9} Research Center for the Early Universe, School of Science, The University of Tokyo, 7-3-1 Hongo, Bunkyo, Tokyo 113-0033, Japan}
\affil{\altaffilmark{10} Astronomy Department, Yale University, 52 Hillhouse Avenue, New Haven, CT 06511, USA}
\affil{\altaffilmark{11} INAF - Osservatorio Astronomico di Padova, Vicolo dell’Osservatorio 5, I-35122 Padova, Italy} 
\affil{\altaffilmark{12} Research School of Astronomy and Astrophysics, Australian National University, Canberra, ACT 2611, Australia}
\affil{\altaffilmark{13} ARC Centre of Excellence for All Sky Astrophysics in 3 Dimensions (ASTRO 3D), Australia}
\affil{\altaffilmark{14} Department of Physics, University of Bath, Claverton Down, Bath, BA2 7AY, UK}

\begin{abstract}
We present 0\farcs2-resolution Atacama Large Millimeter/submillimeter Array observations at 870 $\mu$m in a stellar mass-selected sample of 85 massive ($M_\star>10^{11}~M_\sun$) star-forming galaxies (SFGs) at $z=1.9-2.6$ in the 3D-HST/CANDELS fields of UDS and GOODS-S.
We measure the effective radius of the rest-frame far-infrared (FIR) emission for 62 massive SFGs.
They are distributed over wide ranges of FIR size from $R_\mathrm{e,FIR}=$0.4 kpc to $R_\mathrm{e,FIR}=$6 kpc.  
The effective radius of the FIR emission is smaller by a factor of 2.3$^{+1.9}_{-1.0}$ than the effective radius of the optical emission and by a factor of 1.9$^{+1.9}_{-1.0}$ smaller than the half-mass radius.
Even with taking into account potential extended components, the FIR size would change by $\sim$10\%.
By combining the spatial distributions of the FIR and optical emission, we investigate how galaxies change the effective radius of the optical emission and the stellar mass within a radius of 1 kpc, $M_\mathrm{1kpc}$.
The compact starburst puts most of massive SFGs on the mass--size relation for quiescent galaxies (QGs) at $z\sim2$ within 300 Myr if the current star formation activity and its spatial distribution are maintained.
We also find that within 300 Myr, $\sim$38\% of massive SFGs can reach the central mass of $M_\mathrm{1kpc}=10^{10.5}~M_\sun$, which is around the boundary between massive SFGs and QGs.
These results suggest an outside-in transformation scenario in which a dense core is formed at the center of a more extended disk, likely via dissipative in-disk inflows.
Synchronized observations at ALMA 870 $\mu$m and {\it JWST} 3--4 $\mu$m will explicitly verify this scenario.
\end{abstract}



\section{Introduction}

In the current Universe, the most massive ($M_\star>10^{11}~M_\sun$) galaxies are spheroids or have a prominent bulge in the center \citep[e.g.,][]{2014ARA&A..52..291C} and show little ongoing star formation activity \citep[e.g.,][]{2010ApJ...721..193P}.
One of the ultimate challenges in galaxy formation and evolution is to understand how massive galaxies obtained a bulge-dominated morphology and stopped forming stars.
In this paper, we focus especially on the structural evolution in massive galaxies at $z\sim2$, when the cosmic star formation history peaks \citep[e.g.,][]{2014ARA&A..52..415M}.

Surface brightness profiles of galaxies are well described by S$\acute{\mathrm{e}}$rsic models \citep{1963BAAA....6...41S},

\begin{eqnarray}
I(r)\propto \exp[-b_n(\frac{r}{R_{\mathrm{e}}})^{1/n}-1]
\end{eqnarray}

\noindent
where $R_\mathrm{e}$ is the effective radius, $n$ is the S$\acute{\mathrm{e}}$rsic index ($n=0.5$ for Gaussian, $n=1$ for an exponential profile, $n=4$ for de Vaucouleurs profile), and $b_n$ is an $n$-dependent normalization parameter ensuring that $R_\mathrm{e}$ encloses half the light.
Typical star-forming galaxies (SFGs) are disk-dominated systems ($n \sim 1$) whereas quiescent galaxies (QGs) are bulge-dominated ($n\sim4$).
The correlation between galaxy structure and star formation activity appears to be established at least out to $z \sim 2.5$ \citep[e.g.,][]{2011ApJ...742...96W,2015ApJ...811L..12W}.
Furthermore, central stellar mass surface density and bulge-to-total mass ratio are better correlated to specific star formation rate, color and fraction of quiescent galaxies than total stellar mass, suggesting that it is a good predictor of quiescence \citep{2008ApJ...688..770F,2012ApJ...753..167B,2014ApJ...788...11L,2014ApJ...791...45V}.
Given the result that massive disky QGs are rare, massive SFGs need to change their morphology from disk-dominated to bulge-dominated at some point and quench star formation around that time or soon thereafter.

On the other hand, it is also noted that the observed correlation between galaxy structure and star formation activity does not necessarily imply physical causation. 
\cite{2016ApJ...833....1L} demonstrate, using a simple toy model, that the correlation can naturally arise without any such physical role once the evolution of the size--mass relation for SFGs is taken into account (see also \citealt{2015ApJ...813...23V}). 
It is important to determine how the structural evolution of the galaxy population proceeds by updating the knowledge about spatial distributions of stars and star formation in massive SFGs.

The effective radius of galaxies is another important parameter for understanding the morphological evolution.
Many studies have shown that massive QGs at $z\sim2$ are remarkably compact with $R_\mathrm{e}\sim1$ kpc, which is a factor of 4--5 smaller than the size of local QGs at fixed mass \citep[e.g.,][]{2005ApJ...626..680D,2006MNRAS.373L..36T,2008ApJ...677L...5V,2014ApJ...788...28V}.
Both simulations and observations support that massive compact QGs increase their size through gas-poor minor mergers \citep{2009ApJ...697.1290B,2009ApJ...699L.178N,2012ApJ...746..162N,2014ApJ...780....1W}.
Massive compact SFGs have also been discovered at $z\sim2$ and are naturally expected to be the direct progenitors of compact QGs \citep[e.g.,][]{2013ApJ...765..104B,2014Natur.513..394N,2014ApJ...791...52B,2014ApJ...780....1W,2018ApJ...855...97W}.
The evolutionary link from compact SFGs to compact QGs and eventually giant ellipticals has been extensively studied as mentioned above, while the evolutionary fate of extended SFGs has remained relatively less explored.


In the stellar mass range of $M_\star<10^{11}~M_\sun$, the majority of SFGs at $z=2$ have an extended disk with $R_\mathrm{e}=2-5$ kpc \citep[e.g.,][]{2006Natur.442..786G,2012ApJ...745...85L,2014ApJ...788...28V,2018ApJS..238...21F} and form stars in the more extended disk \citep{2015ApJ...813...23V,2015Sci...348..314T,2016ApJ...828...27N,2019PASJ...71...69S,2020ApJ...892....1W}.
This inside-out growth scenario is also supported by a positive slope in the stellar mass--size relation \citep{2014ApJ...788...28V}.
On the other hand, the inside-out growth alone can not fully explain the high S$\acute{\mathrm{e}}$rsic index ($n\sim4$) in massive QGs or equivalently formation of a dense core.
A combination of the inside-out growth and a declining star formation history can slightly increase the S$\acute{\mathrm{e}}$rsic index of massive SFGs from $n=1$ to $n=1.5-2$ \citep{2016ApJ...833....1L}, but it is not enough.
The question is how and when massive SFGs form a dense core and acquire a radial profile with $n\sim4$.

To answer this question, we need to investigate where stars are formed in SFGs with stellar masses above log($M_\star/M_\sun)=11$, where morphological transformation and quenching are expected to occur \citep[e.g.,][]{2012MNRAS.427.1666B,2012ApJ...753..167B,2014ApJ...788...11L}.
The spatial distribution of stellar mass and ongoing star formation has been well studied over the past decade by observations of H$\alpha$ line and the rest-frame ultraviolet (UV) and optical continuum emission, as mentioned above.
However, in this stellar mass range, 99\% of the total star formation rate (SFR) is obscured by dust, and even H$\alpha$ emission misses 90\%--95\% of star formation \citep{2015ApJ...813...23V,2017ApJ...834..135T,2017ApJ...850..208W}. 
A radial gradient in dust extinction, furthermore, makes it difficult to study the spatial distribution of the total bolometric star formation in massive SFGs using H$\alpha$ and UV \citep{2016ApJ...817L...9N,2018ApJ...859...56T}.


Given that most star formation is obscured by dust, observing the spatially-resolved dust continuum is the best approach to determine where stars truly form in massive SFGs.
For galaxies at $z\sim2.2$, 870 $\mu$m observations sample the dust continuum emission in the rest-frame FIR ($\sim$270 $\mu$m), corresponding to the Rayleigh--Jeans side of modified blackbody radiation with a dust temperature of 25 K.
Total 870 $\mu$m flux densities are commonly used as a tracer of galaxy-integrated gas mass rather than infrared luminosity and SFR \citep{2016ApJ...820...83S} while it is reasonable to use the spatial distribution of 870 $\mu$m continuum emission to locate where the dusty star formation happens in galaxies.
In this paper, we present {\it HST}-resolution observations of 870 $\mu$m continuum emission with the Atacama Large Millimeter/ submillimeter Array (ALMA) to study the spatial distribution of star formation in massive SFGs at $z=2$.
In Section \ref{sec;obs}, we describe the sample selection and ALMA observations. 
We show the results on detection rates of the 870 $\mu$m continuum emission and discuss the sample bias in Section \ref{sec;result}.
We measure the structural parameters of the dust emission through visibility fitting and compare the effective radius in the rest frame FIR with the effective radius in the optical in Section \ref{sec;size}.
Based on the size measurements both at FIR and optical, we discuss structural evolution and quenching in massive galaxies at $z=2$ in Section \ref{sec;discussion}.
Finally, we give a summary in Section \ref{sec;summary}.
We assume a \cite{2003PASP..115..763C} initial mass function (IMF) and adopt cosmological parameters of $H_0$ =70 km s$^{-1}$ Mpc$^{-1}$, $\Omega_{\rm M}$=0.3, and $\Omega_\Lambda$ =0.7.

\begin{figure*}[t]
\begin{center}
\includegraphics[scale=1]{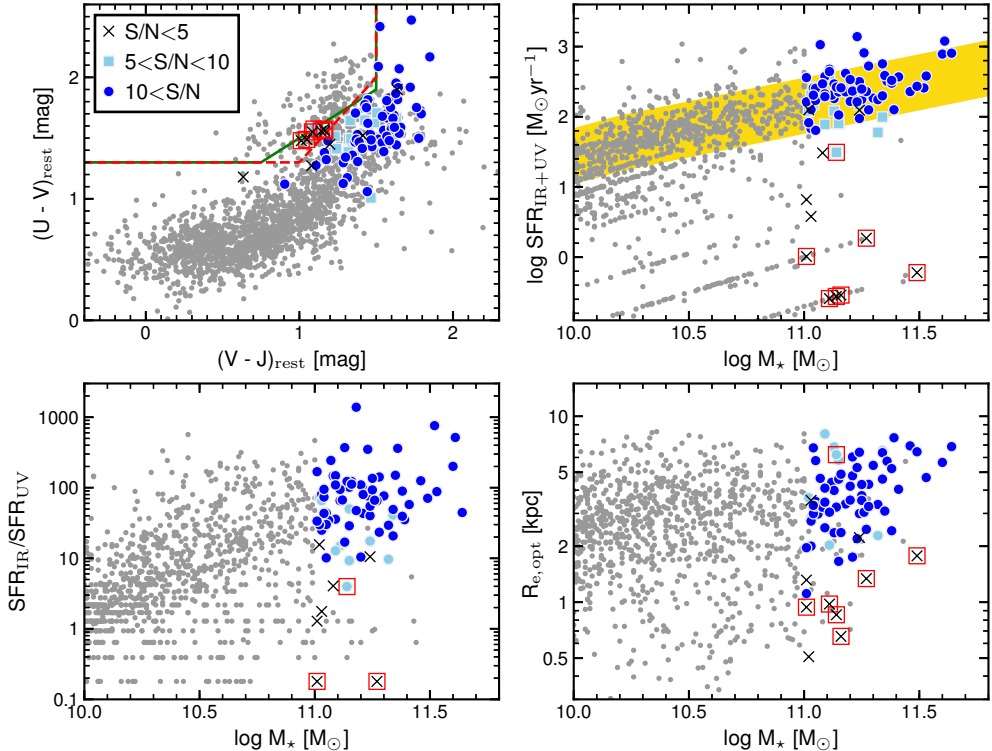}
\end{center}
\caption{Top left: the rest-frame U-V vs. V-J color diagram for our ALMA sample of 85 massive SFGs at $z = 1.9-2.6$. 
Blue circles, light blue squares and black crosses indicate galaxies that are detected at above 10$\sigma$, at $5-10\sigma$ and not detected in the intermediate-resolution 870 $\mu$m images.
Gray dots show all galaxies in the same redshift range, taken from the 3D-HST catalog \citep{2014ApJS..214...24S, 2016ApJS..225...27M}.
A green continuous line shows the criteria to separate between star-forming and quiescent population, used in \cite{2015ApJ...813...23V}, while a red dashed line indicates our modified criteria, defined by Equation (\ref{eq:UVJ}).
Red open squares show 7 galaxies, which are classified as a quiescent population by the modified $UVJ$ criteria.
Top right: stellar mass vs. SFR. Our ALMA targets are mostly located within $\pm0.4$ dex of the star formation main sequence at $2<z<2.5$ (shaded region; a broken power law model in \citealt{2014ApJ...795..104W}). 
Bottom left: stellar mass vs. ratio of obscured to unobscured SFR.
Bottom right: stellar mass vs. effective radius along the semi-major axis in the rest-frame optical \citep{2014ApJ...788...28V}. 
\label{fig;UVJ}
}
\end{figure*}

\section{Sample and observations}\label{sec;obs}

\subsection{Sample selection}\label{sec;selection}

We use the 3D-HST catalog including multi-wavelength photometry from ultraviolet to near-infrared and grism redshifts \citep{2014ApJS..214...24S, 2016ApJS..225...27M} for 94,609 galaxies over 362 arcmin$^2$ of the UDS and GOODS-S fields in the CANDELS survey \citep{2011ApJS..197...35G, 2011ApJS..197...36K}.
Our sample selection for ALMA observations is as follows.

\begin{enumerate}
 \item First, we consider a clean sample of galaxies with reliable photometry including $K$-band (O. Almaini et al. in preparation; \citealt{2008ApJ...682..985W,2010A&A...511A..50R}) by adopting a flag of {\tt use\_phot}=1 (see \citealt{2014ApJS..214...24S}). 
This flag excludes stars, objects close to a bright star and galaxies whose photometric redshift is not derived.
\item Second, we select galaxies in the redshift range of $1.9<z<2.6$. The spectroscopic or grism redshift is used if available and otherwise the photometric redshift is used.
\item Third, we focus on the most massive galaxies with $\log(M_\star/M_\sun)>11$ in this work as this mass range is important for understanding the transformation of galaxy morphology, especially related to bulge formation \citep[e.g.,][]{2012MNRAS.427.1666B,2012ApJ...753..167B,2014ApJ...788...11L}.
Using the {\tt FAST} code \citep{2009ApJ...700..221K}, we perform spectral energy distribution (SED) fitting with stellar population synthesis models of \citet{2003MNRAS.344.1000B} under a \cite{2003PASP..115..763C} initial mass function and the dust attenuation law of \cite{2000ApJ...533..682C} to estimate stellar masses. 
We adopt exponentially declining star formation histories with reasonable constraints on the $e$-folding timescales and stellar ages (see also \citealt{2011ApJ...738..106W}).
\item Finally, we apply the rest-frame $UVJ$ color--color selection to separate SFGs from quiescent galaxies (e.g., \citealt{2007ApJ...655...51W}; \citealt{2011ApJ...735...86W}). 
86 and 26 galaxies fall in the star-forming and the quiescent regime, respectively (Figure \ref{fig;UVJ}). Then, we select only $UVJ$-based SFGs.
\end{enumerate}

We do not apply further selection criteria based on the SFR or optical morphology to ensure we have an unbiased sample of massive SFGs. 
One object (U4-5155) lies between two bright galaxies and is elongated with an effective radius of 5 arcsec in the {\it HST}/F160W-band image \citep{2014ApJ...788...28V}. 
We exclude this galaxy from the sample as it is likely to be magnified by a gravitational lens.
The final sample of 85 SFGs is listed in Table \ref{sourcelist}.
69 and 16 are located in UDS (191 arcmin$^2$) and GOODS-S field (171 arcmin$^2$), respectively.
19 are spectroscopically confirmed with the detection of the H$\alpha$ emission line in the KMOS$^{\mathrm{3D}}$ survey \citep{2015ApJ...799..209W,2019ApJ...886..124W} or the CO emission line by ALMA observations \citep{2017ApJ...841L..25T}.
32 have accurate redshifts ($\Delta z=\pm0.02$) based on the {\it HST} grism spectra \citep{2016ApJS..225...27M} or H$\alpha$ narrow-band imaging \citep{2013ApJ...778..114T}.
For the remaining 34 galaxies, we use photometric redshifts.
12 and 3 targets overlap with the sample of \cite{2017ApJ...834..135T} and \cite{2016ApJ...827L..32B}, respectively.
We note that the 3D-HST catalog uses an {\it HST}/F125W+F140W+F160W combined image for detection \citep{2014ApJS..214...24S}, meaning they are effectively rest-optically selected. 
While the detection limit is much lower mass than $\log(M_\star/M_\sun)=11$, it is possible that some unknown number of massive galaxies could be optically dark and missed \citep[e.g.,][]{2018A&A...620A.152F,2019ApJ...878...73Y}.
Currently, only several optically dark galaxies have been spectroscopically identified and they are all located at $z>3$ \citep{2018ApJ...864...49P,2019Natur.572..211W,2019ApJ...884..154W,2020arXiv200709887U}.
This may be due to that the Balmer break at $z>3$ is redshifted to be at the longer wavelength than the HST/F160W-band, leading to non-detection in the HST images. 
Therefore, optically dark galaxies at $z\sim2$ are likely to have lesser impact on the completeness of our sample.

\subsection{Star formation activity}\label{sec;SFR}

We estimate total SFRs by summing the rate of unobscured star formation measured from the rest-frame 2800 \AA~luminosities, corresponding to UV, and obscured star formation from the total infrared (IR) luminosities following \cite{1998ARA&A..36..189K}, after conversion to a \cite{2003PASP..115..763C} IMF.
We cross-match the 3D-HST, $Spitzer$/MIPS (24 $\mu$m) and $Herschel$/PACS (100 $\mu$m, 160 $\mu$m) catalogs built as described in \cite{2011A&A...532A..90L} and \cite{2013A&A...553A.132M}, with a maximum separation of 1.5\farcs.
The IR luminosities are derived from the single band photometry at, in order of priority, 160 $\mu$m, 100 $\mu$m and 24 $\mu$m \citep{2011ApJ...738..106W}.
Three 160 $\mu$m-detected galaxies (U4-4706, U4-7472 and U4-36568) have a companion within 6\arcsec in the ALMA 870 $\mu$m images (see Section \ref{sec;ALMAobs}). 
We do not use their PACS photometry with a PSF size of FWHM=12\arcsec because it can be contaminated by the emission of the nearby 870 $\mu$ source.
44 and 32 are detected without contamination in the PACS and MIPS band, respectively.
For the remaining 9 galaxies, we use the SFRs from SED fitting.
Typical SFR uncertainties including systematic errors are $\pm$0.2 dex for IR-detected galaxies and $\pm$0.25 dex for the others \citep{2011ApJ...738..106W}. 
In our ALMA sample, the ratio of obscured to unobscured SFR is, on average, SFR$_\mathrm{IR}$/SFR$_\mathrm{UV}$=50, indicating that the dust emission is a dominant probe of the star formation.
Note that our sample is not biased toward a dusty population in the stellar mass range of $\log(M_\star/M_\sun)>11$, as shown in Figure \ref{fig;UVJ}.

As shown in Figure \ref{fig;UVJ}, most of our ALMA targets lie on/around the main sequence of star formation at $z \sim 2$ \citep{2014ApJ...795..104W}. 
The $UVJ$ selection is useful for extracting SFGs without spectroscopic data while it would miss some objects with low-level star formation \citep[e.g.,][]{2017ApJ...841L...6B}.
5 of 26 $UVJ$-selected QGs are detected in the PACS or MIPS band.
It is not straightforward to estimate SFRs of QGs from the IR emission because the IR-based SFRs may be overestimated by contributions from evolved stellar populations \citep{2014ApJ...783L..30U,2019ApJ...877..140L} or time-averaging effects for recently quenched galaxies \citep{2014MNRAS.445.1598H}.
The 5 QGs with IR detection are located near the bottom of the main sequence even if the IR emission is all originating from massive young stars.
They are not included in our sample but it is going to have a minimal impact on our statistical analysis of a representative star-forming population.

\subsection{Optical structure}\label{sec;opt_size}

Deep ${\it HST}$/WFC3 imagining data is available for our sample in the 3D-HST/CANDELS fields.
Fitting of elliptical S$\acute{\mathrm{e}}$rsic models to the two-dimensional light profiles in the F160W-band image gives the structural parameters of the rest-frame optical emission: $R_\mathrm{e,opt}$, $n_\mathrm{opt}$, minor-to-major axis ratio $q_\mathrm{opt}$, and position angle $PA_\mathrm{opt}$ \citep{2012ApJS..203...24V,2014ApJ...788...28V,2014ApJ...788...11L}.
We adopt the effective radius along the semi-major axis to avoid inclination and projection effects, except for Section \ref{sec;discussion} where the circularized effective radius is used.
We do not use 6 objects (U4-4706, U4-7472, U4-16795, U4-23044, GS4-2467 and GS4-5217) for optical size arguments because the best-fit S$\acute{\mathrm{e}}$rsic index reaches the constrained limit ($n = 0.2$ or $n = 8.0$) or the effective radius is unreasonably large with $R_\mathrm{e,opt}>1$\arcsec ($\sim10$ kpc).

\begin{figure}[t]
\begin{center}
\includegraphics[scale=1]{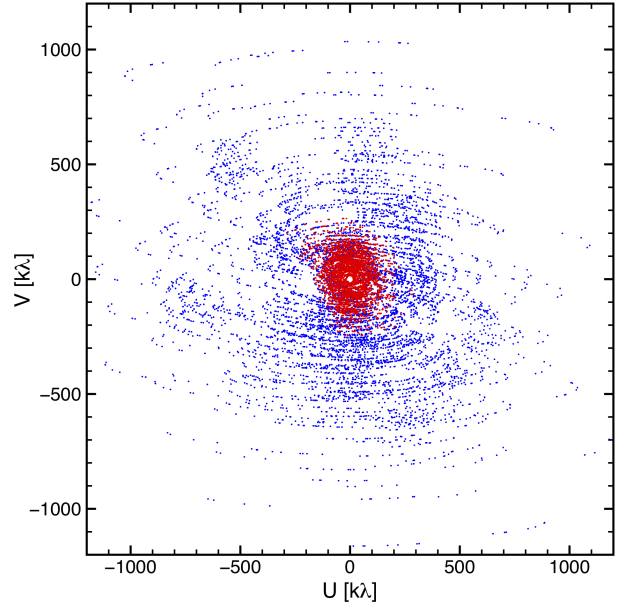}
\end{center}
\caption{An example of the $uv$ coverage in our ALMA observations (U4-190). The extended array data (blue) and the compact one (red) covers the $uv$ range from 14 k$\lambda$ to 1398 k$\lambda$ and from 15 k$\lambda$ to 349 k$\lambda$, respectively.}
\label{fig;uvcoverage}
\end{figure}

\subsection{ALMA observations}\label{sec;ALMAobs}

We have carried out ALMA observations of dust continuum emission in the mass-selected sample of SFGs at $z=2$. 
We use the band 7 receivers with the 64-input correlator in Time Division Mode in a central frequency of 345 GHz ($\sim$870 $\mu$m).
The frequency range available does not cover intermediate-$J$ ($J_\mathrm{up}<7$) CO lines or atomic carbon lines at $z=2$.
We sample a wide range of spatial frequency in the $uv$ plane by using two array configurations: compact array with baseline lengths of 15--300 m and extended array with 15--1260 m (Figure \ref{fig;uvcoverage}).
The extended array data determine the spatial resolution of interferometric images produced through Fourier Transforms while the compact array data are crucial for measuring total fluxes, corresponding to visibility amplitudes in a zero baseline.
If short baseline data were not sufficiently sensitive, the size measurements of 870 $\mu$m emission would be strongly biased by compact components in galaxies.
The maximum recoverable scale of our ALMA observations is 6\farcs7, corresponding to $\sim$55 kpc at $z=2$.
We performed new observations of 84 galaxies with the compact array in May-June 2018 and 66 galaxies with the extended array in September 2018.
The typical on-source time with 43--48 antennas is 4--5 minutes with the compact array and 7--9 minutes with the extended array.
For the other galaxies in our sample, we use ALMA archival data (2012.1.00245.S, 2012.1.00983.S, 2013.1.00205.S, 2013.1.00566.S, 2013.1.00884.S, 2015.1.00242.S, 2016.1.01079.S) to avoid duplicate observations.
We utilize the Common Astronomy Software Application package ({\tt CASA}; \citealt{2007ASPC..376..127M}) for the data calibration. 

We create high-, intermediate-, and low-resolution images by changing the weights of visibilities.
We clean images down to the 1.5$\sigma$ level in a circular mask with a diameter of 2\arcsec\ (4\arcsec\ only for low-resolution images) using the {\tt CASA/tclean} task. 
First, we create high-resolution images by using the compact array data with $uv$-distance of $>180$ k$\lambda$ and all the extended array data and adopting a briggs weighting with a {\tt robust} parameter of +0.5.
The spatial resolution and the noise level is 0\farcs2--0\farcs3 ($\sim$2 kpc) and 60$\pm$7 $\mu$Jy, respectively.
The high-resolution images are used only for visually inspecting the location and the morphology of the dust emission.
Next, we create intermediate-resolution images by using all the compact array data and the extended array data with $uv$-distance of $<180$ k$\lambda$ and adopting {\tt robust}=+2.0.
The spatial resolution and the noise level is 0\farcs8--1\farcs0 ($\sim$7 kpc) and 61$\pm$6 $\mu$Jy, respectively.
The intermediate-resolution images are useful for defining the signal-to-noise ratios (S/N) of 870 $\mu$m continuum emission while they are not sufficient for measuring the total fluxes from the whole galaxies.
We therefore create low-resolution images from all the combined data with {\tt uvtaper} = 4\arcsec\ and {\tt robust}=2.0, resulting in the noise level of 174$\pm$29 $\mu$Jy.
The spatial resolution is 3\farcs3--3\farcs5 ($\sim$ 28 kpc), which should cover whole galaxies at $z=2$.
Note that we do not primarily use these ALMA images for measuring the effective radius of 870 $\mu$m emission (section \ref{sec;vis-fitting}).

\begin{figure}[!t]
\begin{center}
\includegraphics[scale=1]{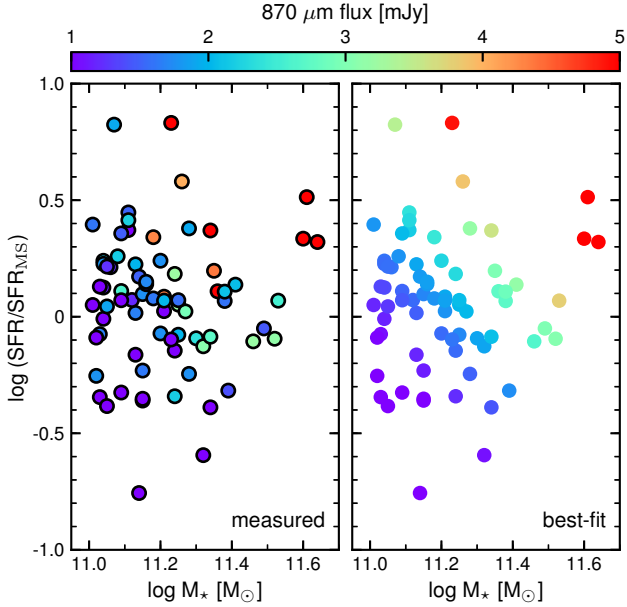}
\end{center}
\caption{870 $\mu$m flux densities for our ALMA sample of 74 galaxies at S/N$>5$, as a function of stellar mass and offset in SFR from the main-sequence.
Left: the peak flux densities measured in the low-resolution images.
Right: the flux densities inferred from the stellar mass and SFR offset using Equation (\ref{eq:flux}). 
}
\label{fig;870flux}
\end{figure}

\section{Results}\label{sec;result}

\subsection{Detections of 870 $\mu$m continuum emission}\label{sec;ALMAresults}

We detect the 870 $\mu$m continuum emission from 74 of 85 galaxies at S/N$>5$, 62 of which have S/N$>10$, in the intermediate-resolution images. 
For 74 detected objects, we measure the total 870 $\mu$m flux density from the low-resolution images.
As U4-7472 and U4-36568 have a companion 870 $\mu$m source within a radius of 1\farcs5, we create low-resolution images after subtracting the companion in the visibility plane to measure the flux densities (section \ref{sec;vis-fitting}).
The median flux density is $S_\mathrm{image}=$1.8 mJy in the range from 0.4 mJy to 6.9 mJy.
Dust continuum emission on the Rayleigh--Jeans regime is commonly used as a tracer of gas mass \citep[e.g.,][]{2015ApJ...800...20G,2016ApJ...820...83S,2018ApJ...853..179T}.
We estimate gas mass in massive SFGs from the 870 $\mu$m flux density by using the calibration of \cite{2016ApJ...820...83S}.
There is a clear trend of increasing flux density in the plane of stellar mass and offset of a galaxy from the star formation main sequence, SFR/SFR$_\mathrm{MS}$, defined by a broken power law model in \citealt{2014ApJ...795..104W}.
More massive, more actively star-forming galaxies are brighter at 870 $\mu$m (Figure \ref{fig;870flux}).
From a linear fitting in this parameter space, we derive equations to predict a 870 $\mu$m flux density and a gas mass of SFGs as,

\begin{eqnarray}
\log S_\mathrm{870,fit} [\mathrm{mJy}]&=&(0.95\pm0.15) \log M_\star \nonumber\\
&&+ (0.52\pm0.08) \log (\frac{\mathrm{SFR}}{\mathrm{SFR}_\mathrm{MS}}) \nonumber\\
&& - (10.41\pm1.7), \label{eq:flux} \\
\log M_\mathrm{gas,fit} [M_\sun]&=&(0.95\pm0.15) \log M_\star \nonumber \\
&&+ (0.53\pm0.08)\log (\frac{\mathrm{SFR}}{\mathrm{SFR}_\mathrm{MS}}) \nonumber \\
&& + (0.33\pm1.7). \label{eq:Mgas}
\end{eqnarray}

\begin{figure}[!t]
\begin{center}
\includegraphics[scale=1]{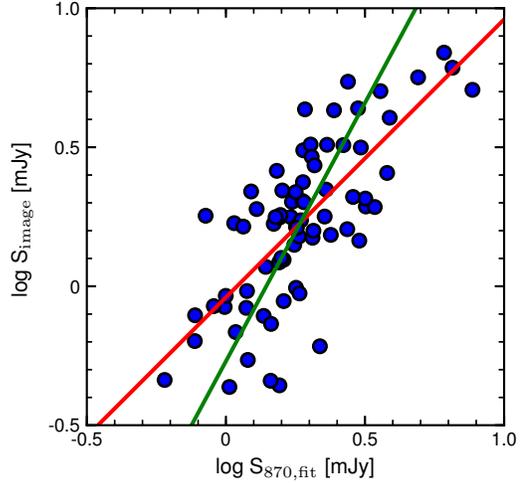}
\end{center}
\caption{A comparison between the predicted and the measured flux densities for our ALMA sample of 74 galaxies at S/N$>5$.
A red line is the best-fit line ($S_\mathrm{image}= 1.00 S_\mathrm{870,fit}-0.04$) from a least squares linear fit when the variable is $S_\mathrm{870,fit}$. 
A green line is the best-fit line ($S_\mathrm{image}= 1.86 S_\mathrm{870,fit}-0.27$) from a least squares linear fit when the variable is $S_\mathrm{image}$. }
\label{fig;870flux_fit}
\end{figure}

\noindent
The dependence on SFR offset is identical to the general scaling relation covering the wide range of galaxy properties \citep{2018ApJ...853..179T} while the mass-dependence is slightly stronger than in the scaling relation (exponent 0.67; \citealt{2018ApJ...853..179T}).
The median and the standard deviation of the difference between the actual measurements and the best-fit values is $\log (S_\mathrm{image}/S_\mathrm{870,fit})=-0.03\pm$0.20 dex. 
On the other hand, there is a systematic error in the predictions by Equation (\ref{eq:flux}): the flux densities are overestimated at $S_\mathrm{image}<$1 mJy and underestimated at $S_\mathrm{image}>$3 mJy (Figure \ref{fig;870flux_fit}).
It is possible to correct for the systematics by applying $S_\mathrm{870,fit,cor}=$1.86$S_\mathrm{870,fit}$-0.27, but the standard deviation of $\log (S_\mathrm{image}/S_\mathrm{870,fit,cor})$ increases to 0.27 dex.
We also note that this prediction is valid only in the stellar mass range of $\log(M_\star/M_\sun)>11$ and in the redshift range of $z=1.9-2.6$.

In addition to the primary targets of massive SFGs at $z\sim2$, we serendipitously detect 19 dust continuum sources at S/N$>$5.0 in the regions within 13.6 arcsec of the phase center, where the primary beam correction factor is less than 0.15 (Table \ref{serendipitous_sourcelist}).
The median redshift is $z=2.7$ and the 870 $\mu$ m flux density ranges from 0.4 mJy to 6.8 mJy after the primary beam correction.
One source (U4-36568b) does not have an optical counterpart and is likely to be a massive SFG at $z>3$ \citep[e.g.,][]{2019Natur.572..211W}.
4 of 19 have a similar redshift to the primary targets with $|\Delta z|<0.1$, suggesting that they are physically associated.
In fact, one (GS4-44920 and GS4-45068) of the four pairs has been spectroscopically identified at $z=2.448-2.450$ \citep{2013A&A...549A..63K}, with a projected separation of 4\arcsec ($\sim32$ kpc).

\subsection{Non-detected objects}\label{sec;nondetection}

For 11 non-detected objects, we give the 5$\sigma$ upper limit of the 870 $\mu$m flux density from the low-resolution images.
6 of them are very close to the boundary of the quiescent regime in the $UVJ$ diagram \citep{2015ApJ...813...23V} and have the lowest SFRs in our sample.
Given that they lie about 2 dex below the main sequence at $z\sim2$ in the $M_\star$--SFR diagram (Figure \ref{fig;UVJ}), they are no longer representative of the star-forming population. 
Therefore, we use the modified $UVJ$ criteria to define a clean sample of massive SFGs at $z\sim2$ as,

\begin{eqnarray}
V-J &<& 1.5, \nonumber \\
U-V &>& 1.3, \nonumber \\
U-V &>& 1.4(V-J)-0.1. \label{eq:UVJ}
\end{eqnarray}

\noindent
The detection rate in the updated SFG sample increases from 87\% to 94\%, indicating it represents an almost mass-complete sample of SFGs down to $\log(M_\star/M_\sun)=11$.

For one galaxy (GS4-40185) classified as QGs in the modified criterion, we detect the 870 $\mu$m continuum emission at S/N=5.5.
This galaxy is located below the main-sequence with $\log$(SFR/SFR$_\mathrm{MS})=-0.76$.
The gas mass fraction is $M_\mathrm{gas}/(M_\star+M_\mathrm{gas})$=15\%.
This is more than twice smaller than the median value (37\%) in 78 SFGs including 5 non-detected sources and is comparable to the gas fraction of a post-starburst galaxy at $z=0.7$ \citep{2017ApJ...846L..14S}.
The less active star formation and the small gas fraction suggest that star formation is being quenched.
Given that other 6 galaxies with similar quiescent $UVJ$ colors are not detected at 870 $\mu$m, 
GS4-40185 is a good candidate for studying the quenching mechanism, that probably works on a short timescale.

\begin{figure*}[!t]
\begin{center}
\includegraphics[scale=1]{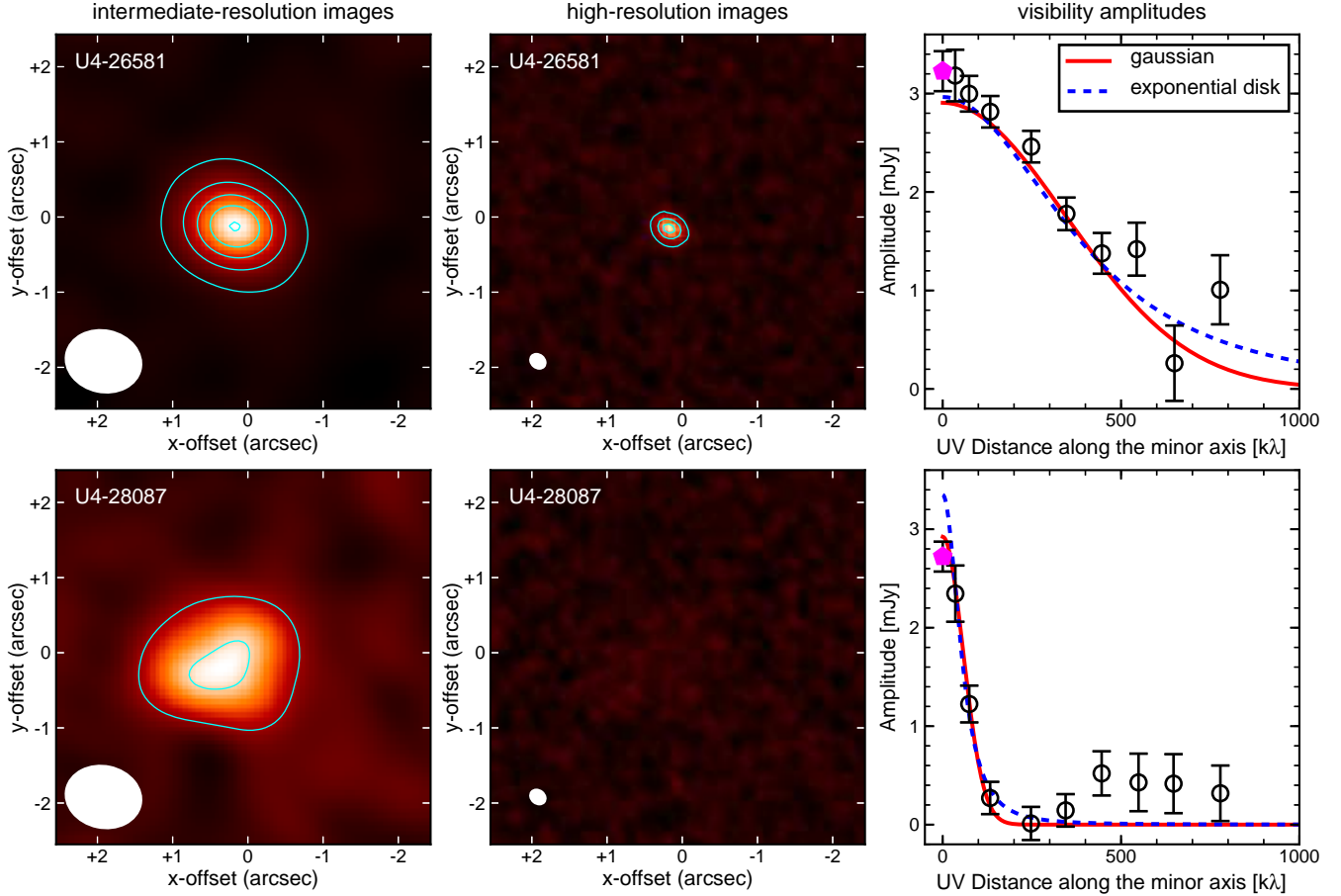}
\end{center}
\caption{ALMA intermediate- and high-resolution images of the 870 $\mu$m continuum emission from U4-26581 (top) and U4-28087 (bottom).
The contours are plotted every 5$\sigma$ in all images. 
Right panels show the visibility amplitudes as a function of $uv$ distance along the minor axis, corresponding to the major axis in the images.
Magenta pentagons show the peak flux density measured in the low-resolution images.
Red solid lines and blue dashed lines indicate the best-fitting model for Gaussian and exponential profile, respectively. 
The fit was done to individual visibilities, not to the averaged ones shown by black circles in this figure. 
}
\label{fig;compact_extended}
\end{figure*}

\section{Size measurements of 870 $\mu$m continuum emission}\label{sec;size}

\subsection{Visibility fitting}\label{sec;vis-fitting}

Size measurements of dust emission usually require high S/N data.
In analysis of images, the appearance and S/N depend largely on both the spatial resolution and the spatial extent of the dust emission.
The spatial resolution is also coupled to the noise level since interferometers sample the Fourier transform of the intensity distribution within a finite $uv$ coverage.
Figure \ref{fig;compact_extended} shows ALMA images of two galaxies with $S_{870}\sim3$ mJy, in which one is compact at 870 $\mu$m and the other is extended.
As the compact object is detected at S/N=29.7 in the high-resolution image, it is possible to accurately measure the size by fitting S$\acute{\mathrm{e}}$rsic models to the image \citep{2015ApJ...807..128S, 2016ApJ...833..103H, 2019MNRAS.490.4956G}.
In contrast, the extended object is completely missed in the high-resolution image.
One easy solution to measure the size is to use the intermediate-resolution image where the emission is detected at S/N=17.1, but inhomogeneous measurements lead to systematic uncertainties.
The non-detection in the high-resolution image roughly means that the dust emission is extended and allows us to constrain the spatial extent.
To fully utilize the information from observations, we measure the size of the 870 $\mu$m continuum emission by looking at how the visibility amplitudes change as a function of spatial frequency \citep{2015ApJ...810..133I, 2015ApJ...807..128S, 2015ApJ...811L...3T}.
The visibility fitting in the $uv$ plane provides homogeneous size measurements, regardless of the compact or extended nature of the sources analyzed (right panels of Figure \ref{fig;compact_extended}).

To derive the structural parameters of the dust emission for 74 massive SFGs with $S/N>5$ in our ALMA sample, we fit elliptical or circular Gaussian models to the observed data by using the {\tt UVMULTIFIT} \citep{2014A&A...563A.136M}.
The details of the fitting process are described in Appendix A and the measured structural parameters are listed in Table \ref{visi-fit}.
Only GS4-40185 (S/N=5.5) shows an extremely small size of FWHM$<$0\farcs01, but this is probably due to the insufficient sensitivity as mentioned above.
We use only 62 massive SFGs with S/N$>$10 for discussion of the size measurements at 870 $\mu$m.

In the GOODS-S field, we also find that there is a systematic offset of $\Delta$R.A=0\farcs10 and $\Delta$Dec.=$-$0\farcs24 in the central positions between {\it HST} and ALMA, which is reported in previous studies \citep{2016ApJ...833...12R,2018A&A...620A.152F}.
In the UDS field, the systematic offset is small, $\Delta$R.A=0\farcs08 and $\Delta$Dec.=$-$0\farcs06.
We correct for these offsets when the ALMA images are visually compared to the {\it HST} images (Figure \ref{fig;HSTimage}).

Although most of the visibility data is well fitted by a Gaussian model, several objects show $>1\sigma$ excesses over the model amplitude at small $u'v'$ and/or large $u'v'$ distances. 
The excess at both small and large $u'v'$ suggest a cuspy profile with an extended tail, corresponding to a higher S$\acute{\mathrm{e}}$rsic index ($n>0.5$).
Surface density profiles of stars and gas in star-forming galaxies are often characterized by an exponential function with $n=1$ \citep{2011ApJ...742...96W,2016ApJ...833..103H}.
We fit the visibility data to exponential disk models by fixing the axis ratio and the position angle to the values derived from Gaussian fitting.
When adopting an exponential disk model, the measured fluxes and effective radii become on average larger by 5\% and 3\%, respectively.

\begin{figure}[t]
\begin{center}
\includegraphics[scale=1]{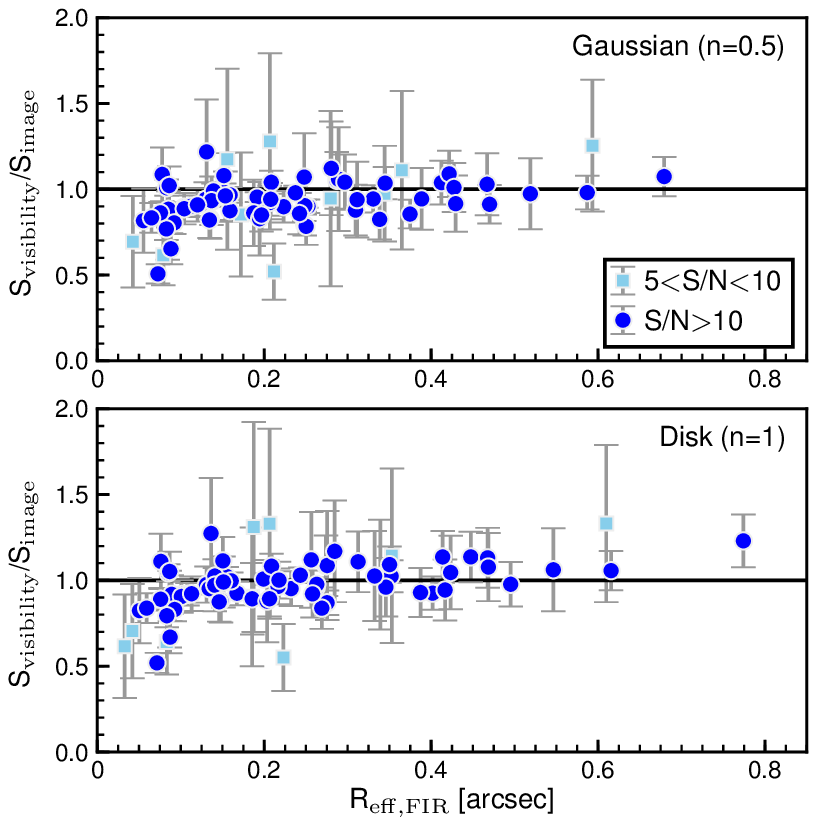}
\end{center}
\caption{Comparisons between visibility-based and image-based flux densities in the low-resolution images as a function of the effective radius of 870 $\mu$m emission. The visibility-based flux densities and sizes are derived from the best-fit model with Gaussian (top) and exponential profile (bottom). 
Blue circles and light blue squares indicate galaxies that are detected at above 10$\sigma$ and at $5-10\sigma$ in the ALMA intermediate-resolution images, respectively.}
\label{fig;flux_comparison}
\end{figure}

\subsection{Contributions of additional extended components}

The visibility amplitudes at the zero baseline inferred from the best-fit models give a total flux density of galaxies.
We compare the visibility-based flux densities for the Gaussian and exponential disk models with the peak flux densities measured in the low-resolution images (Figure \ref{fig;flux_comparison}).
For S/N$>$10 sources, the ratios of the difference between visibility-based and image-based flux densities are on average $S_\mathrm{visibility}/S_\mathrm{image}=$0.93 in the Gaussian and $S_\mathrm{visibility}/S_\mathrm{image}=$0.98 in the disk model, indicating that the dust emission is better characterized by an exponential profile \citep{2016ApJ...833..103H,2019MNRAS.490.4956G}.
We therefore use the effective radius from the disk model in the subsequent sections.
In the disk model, compact objects with $R_\mathrm{e,FIR}<$0\farcs125 show more deficiency of the visibility-based flux densities with $S_\mathrm{visibility}/S_\mathrm{image}=$0.88.
Especially for two compact objects (U4-34138 and U4-34617), the best-fit visibility model includes only 50--70\% of the total flux densities measured in the low-resolution images.
They may have additional components such as an extended halo, off-center clumps or an active galactic nucleus (AGN) or be building cuspy bulges.

For extended objects with $R_\mathrm{e,FIR}>$0\farcs125, the visibility-based flux densities are in good agreement with the image-based one, with $S_\mathrm{visibility}/S_\mathrm{image}=$1.01.
To evaluate the contributions of additional components, we perform a stacking analysis of the model-subtracted visibilities for subsamples of 13 compact objects (excluding U4-34138 and U4-34617) and 47 extended objects, using the {\tt STACKER} tool \citep{2015MNRAS.446.3502L}.
The phase center is individually shifted to the center position of the best-fit model before the stacking. 
We then create low-resolution images from the stacked visibility with {\tt uvtaper} = 4\arcsec and {\tt robust}=2.0, resulting in 3\farcs4~resolution.
The tapered images can in principle recover extended emission even with $R_\mathrm{e,FIR}=$0\farcs5 \citep{2019MNRAS.490.4956G}.
To demonstrate it, we make simulations, which add a Gaussian component with a different flux density to the stacked visibilities and measure the peak flux density in the tapered images. 
The source size is assumed to be $R_\mathrm{e,FIR}=$0\farcs5.
Then, we find that one can recover almost 100\% of the extended emission in the tapered images.

For our ALMA sample, the residual emission we have detected in the tapered images has 220$\pm$57 $\mu$Jy (3.8$\sigma$) for compact objects and 80$\pm$26 $\mu$Jy (3.1$\sigma$) for extended objects.
The contribution to the total flux density is 11\% (4--21\%) for compact objects and 4\% (1--10\%) for extended objects.
We also numerically calculate the effective radius with taking into account the extended component by assuming 2 component models, which consist of the best-fit disk model and an extended gaussian with the stacked flux densities. 
Then, we find that the extended component would change the effective radius by 13\% (4--29\%) for compact objects and by 4\% (1-11\%) for extended objects. 
Therefore, the residual emission including an extended component has less impact on the measurements of the effective radius, except for U4-34138 and U4-34617.

\subsection{Comparison with submillimeter bright galaxies}

\begin{figure}[t]
\begin{center}
\includegraphics[scale=1]{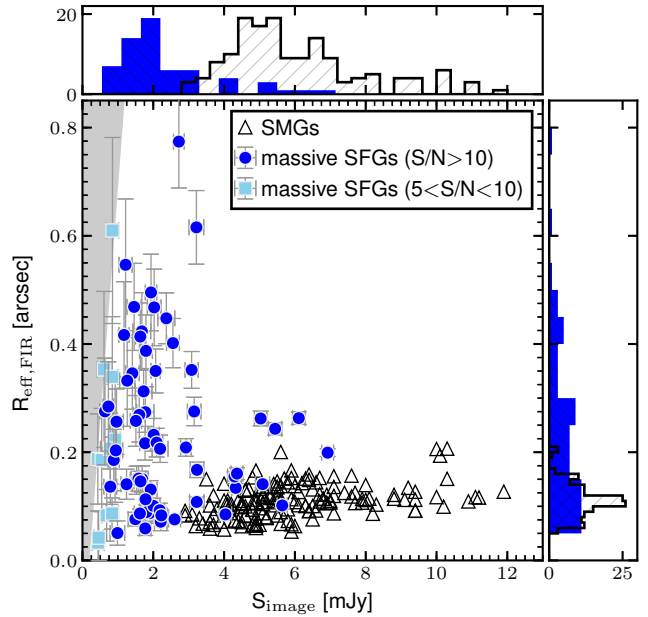}
\end{center}
\caption{870 $\mu$m flux densities and effective radius in massive SFGs at $z=2$ (blue circles and light blue squares) and SMGs at $z=1-4$ (open triangles; \citealt{2019MNRAS.490.4956G}). 
A gray-shaded region shows faint and extended emission that can not be detected at above $5\sigma$ in the simulated observations (Appendix B).
The histogram in the top and right panel is the projected distributions of 870 $\mu$m flux densities and effective radius for massive SFGs (filled blue) and SMGs (open black). 
}
\label{fig;SMG}
\end{figure}

We compare the 870 $\mu$m properties of massive SFGs at $z=2$ with those of submillimeter bright galaxies (SMGs) at $z=1-4$ \citep{2019MNRAS.490.4956G}.
The SMG sample is originally detected as 716 single-dish 850 $\mu$m sources with flux densities of $S_{850}>$3.4 mJy by JCMT survey \citep{2017MNRAS.465.1789G} and is identified as 706 galaxies by ALMA followup observations (AS2UDS; \citealt{2019MNRAS.487.4648S}).
153 of 706 SMGs have robust size measurements from ALMA 0\farcs18-resolution observations \citep{2019MNRAS.490.4956G}.
Our mass-selected sample of SFGs has a median flux density of 1.8 mJy, which is three times fainter than the flux-limited sample of SMGs.
As shown in Figure \ref{fig;SMG}, massive SFGs are distributed over a wide range of the FIR size from $R_\mathrm{e,FIR}=$0\farcs05 to $R_\mathrm{e,FIR}=$0\farcs8 while SMGs are mostly compact with $R_\mathrm{e,FIR}<$0\farcs2.
Other 354 objects in the SMG sample are detected in the same ALMA observations, but their size was not measured due to the low S/N of $<8$. 
The subsample of 153 SMGs is potentially biased to a compact population because the low S/N can be partly caused by extended dust emission as well as faint emission.

On the other hand, the observational bias alone is unlikely to be responsible for the size difference between the two samples.
Even in our ALMA sample, we find that none of 10 bright sources with $S_{870}>$3.4 mJy are extended with $R_\mathrm{e,FIR}>$0\farcs3 while 33\% of the fainter sources are extended.
We also evaluate the completeness in our ALMA observations from the CASA simulations.
Although some faint and extended emission can be missed in our observations, the incompleteness effect is negligible especially at $S_{870}>$1 mJy (a shaded region in Figure \ref{fig;SMG}).
This result confirms the previous results that the bright emission in SMGs originates in their compact core with a high surface density of dust, not in a large disk \citep{2019MNRAS.490.4956G}.
Some physical mechanism that forces a large amount of gas into the central region of galaxies is likely to be at work, causing the compact starburst.

\begin{figure*}[!t]
\begin{center}
\includegraphics[scale=1]{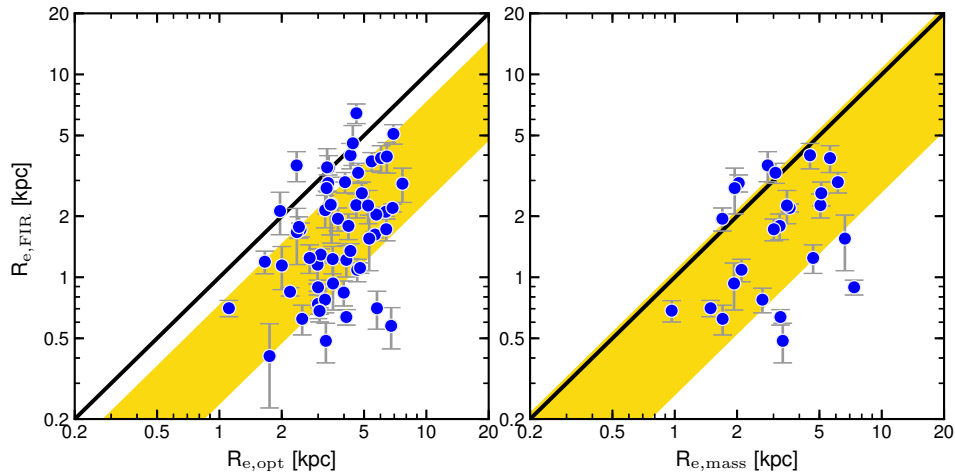}
\end{center}
\caption{Size comparisons between the rest-frame FIR, optical and the stellar mass distribution. 
Yellow-shaded regions show the 16th and 84th percentiles of the size ratios, $R_\mathrm{e,opt}/R_\mathrm{e,FIR}=2.3^{+1.9}_{-1.0}$ and $R_\mathrm{e,mass}/R_\mathrm{e,FIR}=1.9^{+1.9}_{-1.0}$.
}
\label{fig;size_ALMA-HST}
\end{figure*}

\subsection{Comparison of the rest-frame optical and far-infrared size}\label{sec;size_opt-FIR}

Overlay images illustrate that the dust continuum emission is centrally-concentrated while the rest-frame optical emission appears to be extended in the {\it HST}/F160W-band images (Figure \ref{fig;HSTimage} and Figure \ref{fig;HSTimage2} in Appendix A).
As the {\it HST}/F160W-band traces emission in the rest-frame wavelength between $B$-band and $V$-band at $z=1.9-2.6$, it is still affected by dust extinction.
A heavily obscured region does not become bright in the {\it HST} images while it is very bright in the ALMA images.
The ALMA images are not sensitive to low-level, unobscured star formation such as off-center blue clumps \citep[e.g.,][]{2012ApJ...753..114W}.
Nevertheless, dust continuum emission can serve as a primary probe of the spatial distribution of the total star formation in massive SFGs because more than 95\% of the star formation is obscured by dust in the stellar mass range of $\log(M_\star/M_\sun)>11$ (Figure \ref{fig;UVJ}).

4 of 62 massive SFGs with S/N$>10$ do not have robust size measurements in the {\it HST}/F160W-band (Section \ref{sec;opt_size}).
For the remaining 58 massive SFGs in our ALMA sample, the effective radius in the FIR is smaller by a factor of 2.3$^{+1.9}_{-1.0}$ than that in the optical, placing results from previous studies on a more robust footing \citep{2015ApJ...799...81S,2016ApJ...827L..32B,2016ApJ...833...12R,2017ApJ...834..135T,2018ApJ...863...56C,2018A&A...616A.110E,2018ApJ...861....7F,2019ApJ...870..130N,2019ApJ...879...54L,2020A&A...635A.119C}.
The above range is based on the 16th and 84th percentiles of the optical-to-FIR effective radius ratio.
The optical light distribution is commonly used as a probe of stellar mass distribution.
However, the half-light radius is expected to be larger than the half-mass radius especially for old or dusty galaxies where there is a strong radial dependence of mass-to-light ratio \citep[e.g.,][]{2017ApJ...837....2M,2017ApJ...844L...6P,2019ApJ...877..103S}.

For galaxies in the 3D-HST/CANDELS fields, \cite{2014ApJ...788...11L} have constructed a mass-to-light ratio map in the {\it HST}/F160W-band through the resolved SED modeling with the multiband HST data \citep{2012ApJ...753..114W} 
and created a stellar mass map from the {\it HST}/F160W-band map without decreasing the spatial resolution. 
Then, they have fitted single S$\acute{\mathrm{e}}$rsic models to the stellar mass map to derive the half-mass radius.
For our sample, 26 of 58 massive SFGs have measurements of the half-mass radius from \cite{2014ApJ...788...11L}.
The typical uncertainty on the half-mass radius is $\sim$10\% for dusty SFGs \citep{2019ApJ...879...54L}.
Other galaxies are excluded because the best-fit S$\acute{\mathrm{e}}$rsic index reaches the constrained limit ($n$ = 0.2 or $n$ = 8.0) or the effective radius is unreasonably small with $R_\mathrm{e,mass}<$0\farcs05.
The ratio of the half-mass to the half-light radius is $R_\mathrm{e,mass}/R_\mathrm{e,opt}=0.8^{+0.5}_{-0.3}$, which can be interpreted as that the optical continuum emission is indeed affected by strong dust extinction in the center of galaxies.
However, the effective radius in the FIR is still by a factor of 1.9$^{+1.9}_{-0.9}$ smaller than the half-mass radius (Figure \ref{fig;size_ALMA-HST}).
These results suggest that many massive SFGs intensively form stars in a compact central region, embedded in a more extended disk probed by the optical continuum emission.

We also compare the FIR size with the H$\alpha$ size.
8 of 58 massive SFGs have measurements of the effective radius of the H$\alpha$ emission from the KMOS$^{\mathrm{3D}}$ survey \citep{2019ApJ...886..124W,2020ApJ...892....1W}.
The median ratio between the H$\alpha$ and the FIR effective radius is $R_\mathrm{e,H\alpha}/R_\mathrm{e,FIR}=$2.3 though both trace star formation.
The more heavily obscured regions is unlikely to be probed by the rest-frame UV and H$\alpha$ emission \citep{2020A&A...635A.119C}.

We caution that conversion from 870 $\mu$m flux to dust-obscured SFR depends on dust properties such as dust temperature and optical depth.
When dust temperature is higher in a galaxy center, the spatial distribution of infrared luminosity becomes more compact than that of dust continuum emission.
In contrast, the spatial distribution of dust mass becomes more extended than that of dust continuum emission.
This can explain the result from previous studies that the effective radius of intermediate-$J$ ($J_\mathrm{up}=3$ and 4) CO emission is larger than that of dust continuum emission in massive SFGs and SMGs \citep{2017ApJ...846..108C, 2017ApJ...841L..25T, 2018ApJ...863...56C, 2019ApJ...876....1T}.
Simulations also show that the spatial distribution of dust emission is more compact than the total gas component, but has a similar spatial distribution as the cold, dense gas \citep{2019MNRAS.488.1779C}. 
An optical depth effect on dust continuum emission could also lead to a smaller size of star formation because emission at shorter wavelengths (tracing an infrared luminosity) is preferentially suppressed in a galaxy center. 
Therefore, the distribution of dust-obscured SFR could be even more compact than measured from the 870 $\mu$m continuum observations, supporting our result that massive SFGs form stars in a compact region.

\begin{figure*}[!t]
\begin{center}
\includegraphics[scale=1]{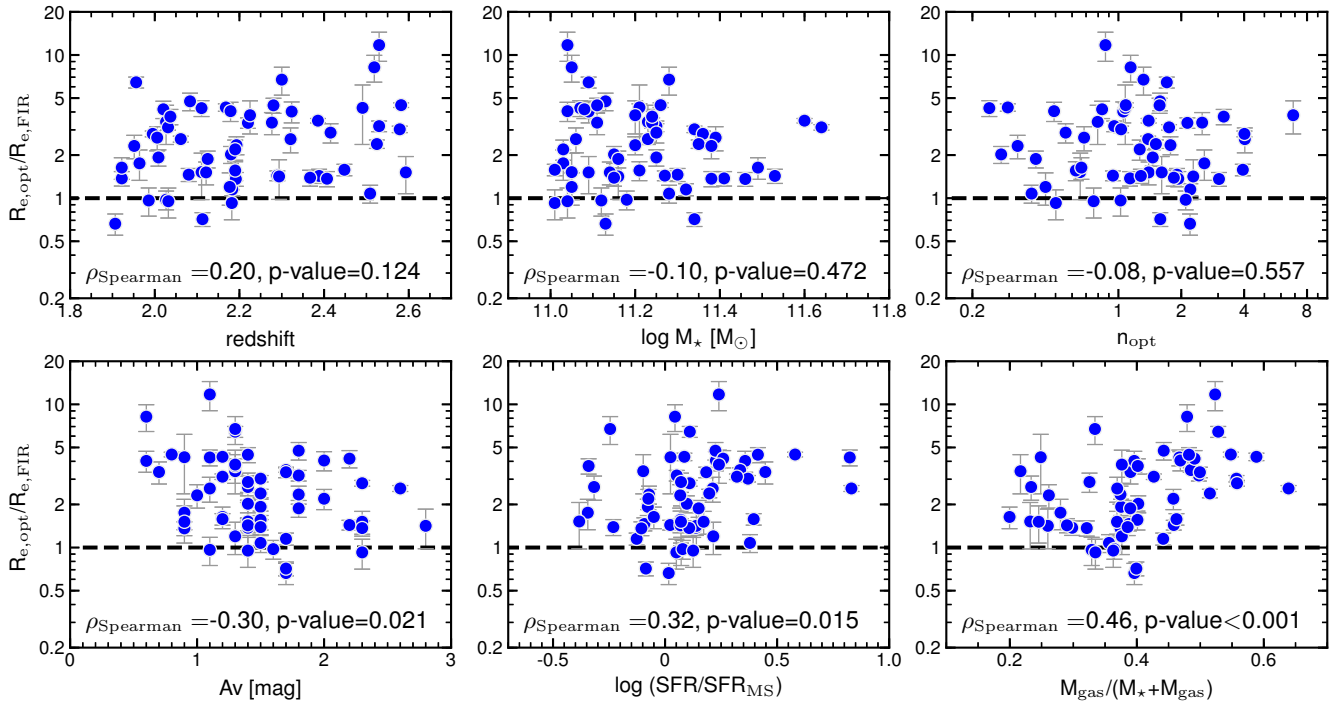}
\end{center}
\caption{Optical-to-FIR effective radius ratio vs. redshift (top left), stellar mass (top center), S$\acute{\mathrm{e}}$rsic index (top right), dust extinction (bottom left), offset from the star formation main-sequence (bottom center), and gas mass fraction (bottom right).
}
\label{fig;optsize-ratio}
\end{figure*}

\subsection{Correlations between the size ratio and other galaxy parameters}

In the previous section, we found that the effective radius of the FIR emission is smaller than that of the optical emission in massive SFGs.
We investigate how the optical-to-FIR effective radius ratio is correlated with other galaxy parameters, including redshift, stellar mass, S$\acute{\mathrm{e}}$rsic index in the optical, dust extinction $A_V$, offset from the star formation main-sequence, and gas mass fraction (Figure \ref{fig;optsize-ratio}).
Then, we compute the Spearman rank correlation coefficient and the $p$-value to test for non-correlation.
When the $p$-value is very close to 0, we can reject the null hypothesis that the observed correlation is zero.
In regard to the redshift, stellar mass, and S$\acute{\mathrm{e}}$rsic index, there is no significant correlation with the size ratio.
The optical-to-FIR effective radius ratio is weakly anticorrelated with the dust extinction, derived from SED fitting to the galaxy-integrated photometry from the rest-frame UV to near-IR, probed by observed optical to mid-IR fluxes.
This is the opposite of what is expected because strong dust extinction in galaxy centers can make the optical effective radius larger.
At least, we do not find evidence that the optical-to-FIR effective radius ratio is larger due to dust extinction.
Also, the ratio is positively correlated with the gas mass fraction and the offset from the star formation main-sequence.
The Spearman rank correlation coefficient and the $p$-value is $\rho_\mathrm{spearman}=$0.46 and $p<0.001$ for gas mass fraction and $\rho_\mathrm{spearman}=$0.32 and $p=0.015$ for main-sequence offset, respectively.

The more centrally-concentrated star formation than stars indicates that gas is transported from the outer disk into the galaxy center.
The fact that more gas-rich, more actively star-forming galaxies tend to have larger $R_\mathrm{e,opt}/R_\mathrm{e,FIR}$ suggests the possibility that the more compact dust emission is related to dissipative processes such as wet compaction.
A dissipative concentration of gas ({\it wet compaction}) is seen in simulations of gas-rich galaxies at high-redshift \citep[e.g.,][]{1999ApJ...514...77N,2004A&A...413..547I,2014ApJ...780...57B,2015MNRAS.450.2327Z,2016MNRAS.457.2790T}.
Wet compaction is triggered by gas-rich minor/major mergers, counter-rotating streams, and radial gas inflows driven by violent disk instability \citep[e.g.,][]{2014MNRAS.438.1870D,2014MNRAS.438.1552F}.
There is also observational evidence of a gas component falling into an SMG with a compact core from a direction perpendicular to the disk rotation \citep{2020ApJ...889..141T}. 
The onset of wet compaction is likely to occur when gas inflows dominate star formation and outflows \citep{2016MNRAS.457.2790T}.
It is therefore expected that the incidence of compaction depends on star-forming activity and gas mass fraction in galaxies.
When gas falls into the galaxy center to feed a core, it has to loose angular momentum. 
The extended H$\alpha$ disk (section \ref{sec;size_opt-FIR}) may be formed by the released angular momentum. 

\section{Discussion}\label{sec;discussion}

Since we derive the spatial distribution of stellar continuum emission from the {\it HST}/F160W data and the distribution of dusty star formation from the ALMA data, 
we combine the two to predict how massive SFGs will change their effective radius and their central stellar mass within a radius of 1kpc, $M_\mathrm{1kpc}$, by in-situ star formation.
We discuss the current and future structure for 58 massive SFGs with robust measurements of the structural parameters in both {\it HST}/F160W and ALMA data.
Unlike the previous sections, we adopt the circularized effective radii of the optical and the FIR emission to treat them fairly since the axis ratios and the position angles are not exactly the same.

\subsection{Spatially-resolved stellar mass and SFR}\label{sec;resolved_MS}

Before investigating the structural evolution of massive SFGs at $z\sim2$, we look at their spatially-resolved stellar mass and SFR at the observed epoch.
The surface stellar mass profile and the surface SFR profile are derived from the best-fitting S$\acute{\mathrm{e}}$rsic profile in the {\it HST}/F160W and ALMA/870 $\mu$m data, respectively.
Following previous studies \citep{2009ApJ...697.1290B,2019ApJ...880...57M}, we deproject the two-dimensional S$\acute{\mathrm{e}}$rsic profile using an Abel transform (see also Appendix C).
We scale the integral of the inferred three-dimensional profile at {\it HST}/F160W to the total stellar mass and that at ALMA 870 $\mu$m to the total SFR to compute the central mass ($M_\mathrm{1kpc}$) and the central SFR within a radius of 1kpc (SFR$_\mathrm{1kpc}$).
Once the central mass and SFR are given, it is possible to estimate the stellar mass and SFR in the outer disk with $R>1$ kpc by subtracting $M_\mathrm{1kpc}$ and SFR$_\mathrm{1kpc}$ from the total stellar mass and the total SFR.

\begin{figure}[!t]
\begin{center}
\includegraphics[scale=1]{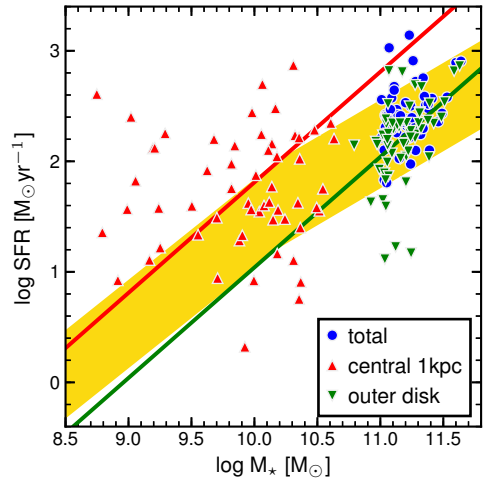}
\end{center}
\caption{The stellar mass vs. SFR for the central 1 kpc regions (red triangles) and the outer disks (green triangles) of massive SFGs. 
Red and green lines correspond to the median sSFR for the central 1 kpc regions (log sSFR [Gyr$^{-1}$]=0.83) and the outer disks (log sSFR [Gyr$^{-1}$]=0.04), respectively.
A yellow shaded region shows the $\pm0.4$ dex range of the star formation main sequence at $2<z<2.5$ \citep{2014ApJ...795..104W}. 
We also show the total stellar mass and total SFR by blue circles.
}
\label{fig;MS1kpc}
\end{figure}

Figure \ref{fig;MS1kpc} shows a spatially-resolved version of the mass--SFR plane for our ALMA sample.
The outer disks are responsible for the total stellar mass while the SFR in the central 1kpc regions is comparable to that in the outer disks.
This leads to a significant difference in the star formation activity.
About half of the central 1kpc regions are located above the star forming main-sequence defined by galaxy-integrated properties while the outer disks mostly lie on the main-sequence.
The median value and the standard deviation of the specific SFRs, defined as sSFR=SFR/$M_\star$, is log sSFR [Gyr$^{-1}$]=0.83$\pm$0.71 for the central 1kpc region and log sSFR [Gyr$^{-1}$]=0.04$\pm$0.33 for the outer disk, indicating that the sSFR is higher in the galaxy center than in the outer disk.

The centrally-rising sSFR profiles seem to be at odds with previous studies that have applied dust gradient corrections to translate H$\alpha$ profiles to SFR profiles, finding that dust-corrected sSFR profiles are centrally dipping in massive SFGs at $z\sim2$ \citep{2015Sci...348..314T,2018ApJ...859...56T,2019PASJ...71...69S}.
This difference may be due to underestimates of the central dust extinction and/or the different sample selection.
\cite{2016ApJ...817L...9N} find significant dust attenuation toward centers of massive SFGs through the Balmer decrement although it is not very deep into dust-embedded regions as can be probed by FIR emission.
In our ALMA observations, it was difficult to derive the spatial distributions of star formation for less dusty SFGs near the bottom of the star-forming main sequence, corresponding to S/N$<$10 sources.
As shown by \cite{2018ApJ...859...56T}, massive less dusty SFGs are likely to have centrally-dipping sSFR profiles.
Dust extinction corrections based on the rest-frame UV color, as applied by \cite{2018ApJ...859...56T}, may be prone to saturation effects in the case of a mixture between obscuring dust clouds and the emitting sources.
Also, many of massive SFGs are very faint and fuzzy in the {\it HST}/F814W-band images, making it difficult to determine the rest-frame UV color at $z\sim2$ in the first place.
Therefore, it would be necessary to use the appropriate indicators, depending on the galaxy properties: submillimeter continuum for massive, dusty SFGs on/around the star-forming main sequence and H$\alpha$ line emission for less dusty SFGs.

\subsection{Structural evolution in massive star-forming galaxies}\label{sec;structural_evolution}

\begin{figure*}[!t]
\begin{center}
\includegraphics[scale=1]{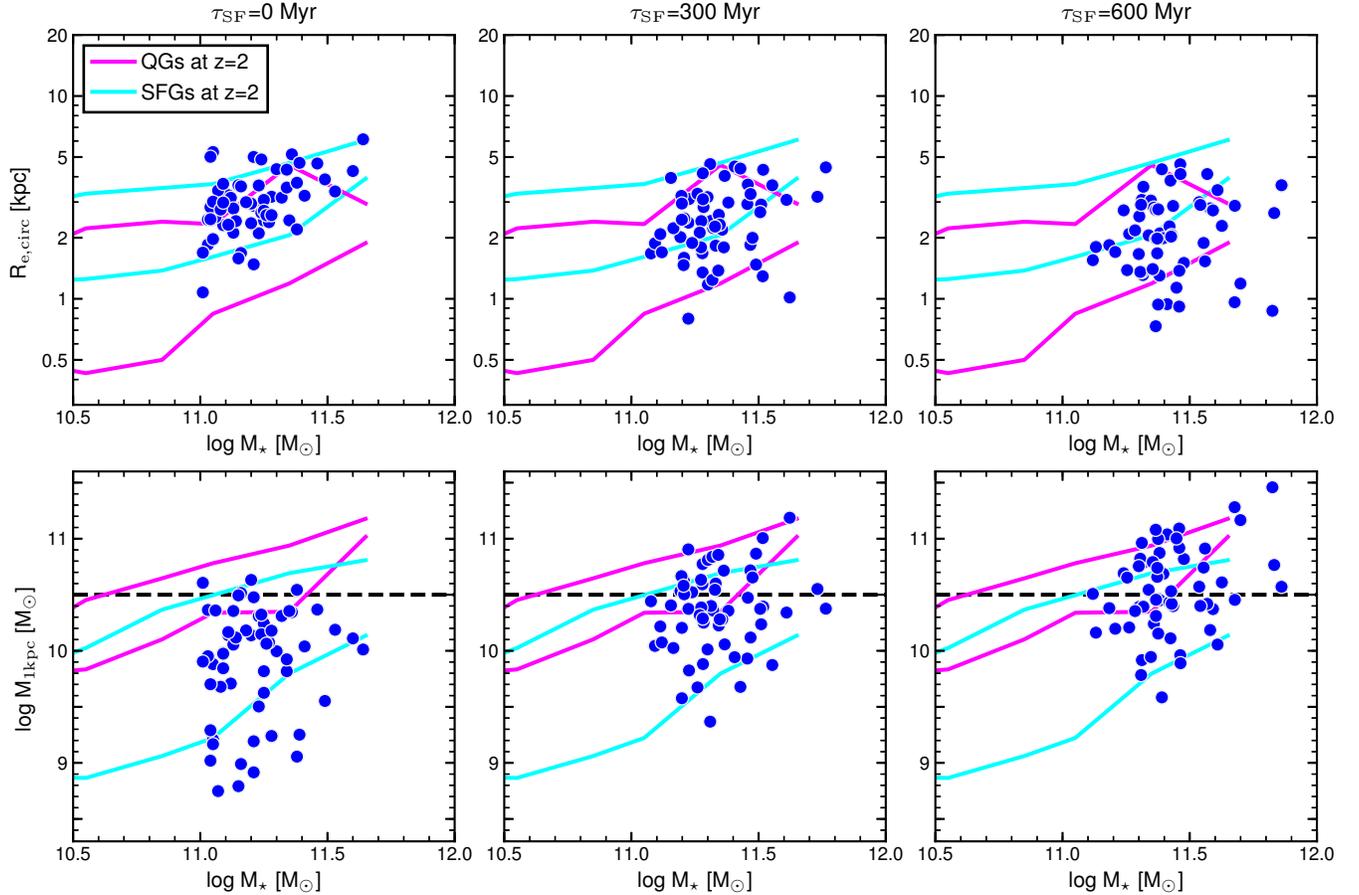}
\end{center}
\caption{
The effective radius of the optical emission (top) and the central 1 kpc mass (bottom) as a function of stellar mass for 58 massive SFGs in our ALMA sample. 
We show the observed values at $\tau_\mathrm{SF}=0$ (left) and the inferred values at $\tau_\mathrm{SF}=300$ Myr (center) and $\tau_\mathrm{SF}=600$ Myr (right), which are based on the spatial distribution of the 870 $\mu$m continuum emission (Section \ref{sec;structural_evolution}).
Cyan and magenta lines indicate the 16th and 84th percentiles in each bin of stellar mass for SFGs and QGs at $z=1.9-2.6$ in the combined sample of the CANDELS/3D-HST survey \citep{2014ApJ...788...28V,2014ApJS..214...24S, 2016ApJS..225...27M} and the COSMOS-DASH survey \citep{2019ApJ...880...57M}. 
Black dashed lines indicate $\log (M_\mathrm{1kpc}/M_\sun)=10.5$, which defines a dense core in this work.
}
\label{fig;mass-size_evolution}
\end{figure*}

Next, we investigate the time evolution of the effective radius and the central mass assuming that the current level of star formation and its spatial distribution are the same for several hundred megayears.
The mass of newly formed stars is estimated to be

\begin{eqnarray}
M_{_\star ,\mathrm{new}}=\mathrm{SFR}\times \tau_\mathrm{SF} \times \beta \label{eq;newstar}
\end{eqnarray}

\noindent
where $\tau_\mathrm{SF}$ is the timescale for star formation and $\beta$ is the mass loss parameter due to supernova explosions and stellar winds. 
We adopt $\beta=0.6$ appropriate for a \cite{2003PASP..115..763C} IMF.
Figure \ref{fig;mass-size_evolution} shows the circularized effective radius and the central mass inside a radius of 1 kpc as a function of stellar mass at $\tau_\mathrm{SF}=0$ Myr, 300 Myr and 600 Myr.
We overplot the 16th and 84th percentiles in each mass bin for SFGs and QGs at $z=1.9-2.6$ in the combined sample of the CANDELS/3D-HST survey \citep{2014ApJ...788...28V,2014ApJS..214...24S, 2016ApJS..225...27M} and the COSMOS-DASH survey \citep{2019ApJ...880...57M}. 

We find that by 300 Myr with the compact starburst, most of massive SFGs lie on the mass-size relation for QGs (Figure \ref{fig;mass-size_evolution}).
This is one of the most important results of this work.
In our ALMA sample of 58 massive SFGs, the effective radius is decreased, on average, by 14\% in 300 Myr while the stellar mass is increased by 28\%.
The averaged evolution in the mass--size plane is approximated by $\Delta \log R_\mathrm{e}=-0.6\Delta \log M_\star$.
The negative slope appears to contradict the context of inside-out growth where galaxies form stars in a more extended disk and gradually increase their size with $\Delta \log R_\mathrm{e}\sim0.3\Delta \log M_\star$ \citep{2015ApJ...813...23V}.
Our result is preferred for explaining the morphological transformation of massive SFGs, at least in terms of effective radius.


We also infer the effective radius of the massive SFGs at higher redshift by putting $\tau_\mathrm{SF}=$ $-$300 Myr in Equation (\ref{eq;newstar}).
Many of them are found to deviate from the observed mass--size relation and become too large especially in the stellar mass range of log($M_\star/M_\sun)<11$, suggesting that very little time has passed since the compact starburst occurred.
Extended star formation in disks is required for explaining the size evolution of SFGs, but it does not transform the morphology of galaxies from disk-dominated to bulge-dominated.
The compact starburst naturally produces a cuspy profile, characterized by a high S$\acute{\mathrm{e}}$rsic index.
This outside-in transformation should happen after the inside-out growth of galaxy disks.
Observations of massive, $\log(M_\star/M_\sun)>11$, early-type galaxies at $z\sim0$ show that the mass-weighted age in the galaxy center is younger than that in the outer disks \citep{2017MNRAS.466.4731G}, supporting the outside-in transformation.

On the other hand, the fact that going 300 Myr back in time brings many galaxies outside the mass--size relation suggests not only that the compact starburst can not have been going on for a long time, but also that it will not go on for a long time after the current time of the observations either.
As it is implausible to find all observed systems in the first fraction of their compact starburst phase, it may not be reasonable to extrapolate the current distributions of star formation into the future beyond 300 Myr.
This is also supported by the result that some of massive SFGs become more compact, $R_\mathrm{e}<1$ kpc, than massive QGs at $\tau_\mathrm{SF}$=600 Myr.

Based on the current distribution of star formation, we also find that 22 ($\sim$38\%) of 58 massive SFGs have a dense core with $\log (M_\mathrm{1kpc}/M_\sun)>10.5$ at $\tau_\mathrm{SF}=$ 300 Myr and additional 8 (a total of 30, $\sim$52\%) reach this threshold at $\tau_\mathrm{SF}=$ 600 Myr.
Further star formation will allow the remaining SFGs to form a dense core in about 2 Gyr, but it is unlikely to happen as mentioned above.
The compact starburst probed by ALMA observations makes $\sim$40\% of massive SFGs form a dense core with $\log(M_\mathrm{1kpc}/M_\sun)>10.5$
whereas $\sim$75\% of massive QGs with log$(M_\star/M_\sun)>$11 reach this threshold (Figure \ref{fig;mass-size_evolution}).
Despite the fact that most of our sample has highly compact dust emission, more than a half of massive SFGs do not form enough stars in this episode to reach the central mass density of $\log(M_\mathrm{1kpc}/M_\sun)>10.5$.
They need to make their distribution of star formation more compact to form a dense core if galaxies transform their morphology before quenching.
An alternative view is that massive SFGs evolve into disky QGs without a dense core, although measurements of S$\acute{\mathrm{e}}$rsic index indicate that such cases are infrequent \citep[e.g.,][]{2011ApJ...742...96W,2015ApJ...811L..12W}. 
Recent studies have found that three gravitationally lensed QGs at $z\sim2$ are rotationally supported and one of them has a perfect exponential disk with no evidence for a dense core \citep{2017Natur.546..510T,2018ApJ...862..125N,2018ApJ...862..126N}.
Some of massive SFGs may be the immediate progenitors of disky QGs at $z\sim2$, but it could be a minor pathway as indicated by the fact that 75\% of massive QGs have a dense core.

\section{Summary}\label{sec;summary}

Using ALMA, we have made high-resolution observations of 870 $\mu$m continuum emission for the mass-selected sample of 85 massive, $\log(M_\star/M_\sun)>11$, SFGs at $z=1.9-2.6$ in the CANDELS/3D-HST fields of UDS and GOODS-S.
We have detected the dust continuum emission from 74 galaxies and robustly measured the effective radius for 62 galaxies.
58 of them have robust size measurements in the {\it HST}/F160W band.
Combining the distribution of star formation from the ALMA data with that of stellar continuum from the {\it HST} data, we have investigated the structural evolution in massive SFGs at $z=2$. 

\begin{enumerate}

\item The 870 $\mu$m flux density ranges from 0.4 mJy to 6.9 mJy and increases as a function of stellar mass and offset from the star-forming main sequence.
As the gas mass can be estimated from the 870 $\mu$m flux density, we confirmed the trend that more massive, more active SFGs have a larger gas reservoir.
6 of 11 non-detected sources are located near the boundary between star-forming and quiescent population in the $UVJ$ diagram.
Once we adopt the slightly modified boundary, the detection rate is 94\% (73/78), indicating that our targets represent a nearly mass-complete sample of SFGs.

\item The effective radius of the 870 $\mu$m emission widely ranges from 0.4 kpc to 6 kpc. Many massive SFGs are more extended in the FIR than bright SMGs with $S_{870}>3.4$ mJy at $z=1-4$.
The significant difference of the FIR size between the mass-selected and the flux-selected sample can be partly caused by the observational bias, where high-resolution ALMA images are not sensitive to extended emission.
On the other hand, there is a lack of bright extended sources with $S_{870}>3.4$ mJy and $R_\mathrm{e,FIR}>$0\farcs3 even in our ALMA sample of massive SFGs, indicating that the bright submillimeter emission originates in not their large disk but their compact core.

\item The effective radius in the FIR is smaller by a factor of 2.3$^{+1.9}_{-1.0}$ than the effective radius in the optical and smaller by a factor of 1.9$^{+1.9}_{-1.0}$ than the half-mass radius.
The compact dust emission suggests that many massive SFGs intensively form stars in the central 1 kpc region, embedded in a more extended disk probed by the optical continuum emission.

\item We found that the FIR size is weakly anticorrelated with the gas mass fraction: gas-rich galaxies tend to be associated with compact dust emission.
This evidence is consistent with dissipative concentration of gas towards the center of gas-rich galaxies via efficient radial inflows

\item We derive the stellar masses and the SFRs separately in the central 1 kpc region and in the outer disk by exploiting the best-fit S$\acute{\mathrm{e}}$rsic profiles in the {\it HST}/F160W and ALMA data. 
The central 1 kpc regions are located above the star formation main-sequence and have higher sSFRs than those in the outer disk, indicating the centrally-rising sSFR profiles.
This result supports an outside-in transformation scenario in which a dense core is formed at the center of a more extended disk, likely via dissipative in-disk inflows.

\item The compact starburst could put most of massive SFGs on the mass--size relation for QGs at $z\sim2$ within 300 Myr if the current star formation activity and its spatial distribution are maintained.
The averaged evolution in the mass--size plane is approximated by $\Delta \log R_\mathrm{e}=-0.3\Delta \log M_\star$.
We also found that 38\% of massive SFGs can form a dense core with $\log(M_\mathrm{1kpc}/M_\sun)=10.5$ within 300 Myr.
Some of the remaining ones may need to make their distribution of star formation more compact by further dissipative processes.

\end{enumerate}

The compact dust emission in our ALMA sample appears to be inconsistent with previous results that the H$\alpha$ emission is more extended than the stellar continuum emission in less massive, less dusty SFGs at $z\sim2$ \citep{2019PASJ...71...69S,2020ApJ...892....1W}.
Note that our targets are all massive SFGs with $\log(M_\star/M_\sun)>11$, where most of star formation is obscured by dust.
In the stellar mass range of $\log(M_\star/M_\sun)<11$, the distribution of dust-obscured star formation is poorly investigated.
Deep intermediate (0\farcs5)-resolution ALMA observations, rather than high-resolution ones, are the best for confirming if less massive SFGs form stars in a more extended disk than the spatial extent of previously formed stars and grow from inside out.

One caveat of our study is that we may overestimate the half-mass radius of massive SFGs due to strong dust extinction in the galaxy centers because the current spatially-resolved SED modeling is based on data that does not extend redwards of rest-frame 5000 \AA, provided by HST/F160W-band observations.
If this is true, the half-mass radius could be as compact as the 870 $\mu$m emission and the S$\acute{\mathrm{e}}$rsic index could already reach $n=4$. 
In that case, massive SFGs do not have to transform their morphology in term of both size and radial profile (core mass). 
At this moment, there is no evidence supporting the possibility that the half-mass radius is overestimated by a factor of two.
It is definitely important to verify if the stellar mass distribution is surely more extended than the dust in massive SFGs. 
They are sufficiently bright at 3--4 $\mu$m, where the emission is less affected by dust extinction, with an AB magnitude of 20--21. 
High-resolution 3--4 $\mu$m observations with Near Infrared Camera on the {\it James Webb Space Telescope (JWST)} will allow us to directly probe the stellar mass distribution.
The ALMA-{\it JWST} synergetic observations will provide a definitive answer in the the structural evolution of massive galaxies at the peak of cosmic star formation history.

\vspace{0.5cm}

We are very grateful to the referee for constructive suggestions to improve the paper.
We thank Pieter van Dokkum for comments on the first draft of the paper.
This paper makes use of the following ALMA data: ADS/JAO.ALMA\#2017.1.01027.S (primary), 2012.1.00245.S, 2012.1.00983.S, 2013.1.00205.S, 2013.1.00566.S, 2013.1.00884.S, 2015.1.00242.S, 2016.1.01079.S. 
ALMA is a partnership of ESO (representing its member states), NSF (USA) and NINS (Japan), together with NRC (Canada) and NSC and ASIAA (Taiwan) and KASI (Republic of Korea), in cooperation with the Republic of Chile. The Joint ALMA Observatory is operated by ESO, AUI/NRAO and NAOJ.
K.T. acknowledges support by JSPS KAKENHI Grant Number JP20K14526. 
E.W. acknowledges support by the Australian Research Council Centre of Excellence for All Sky Astrophysics in 3 Dimensions (ASTRO 3D), through project number CE170100013.
Data analysis was in part carried out on the common-use data analysis computer system at the Astronomy Data Center (ADC) of the National Astronomical Observatory of Japan.
This research made use of {\tt UVMULTIFIT} and {\tt stacker}, software tools developed by the Nordic ALMA Regional Center. The Nordic ARC node is funded through Swedish Research Council grant No 2017-00648.

\section*{Appendix A}
\section*{ALMA images and catalogs}

\begin{figure*}[t]
\begin{center}
\includegraphics[scale=1]{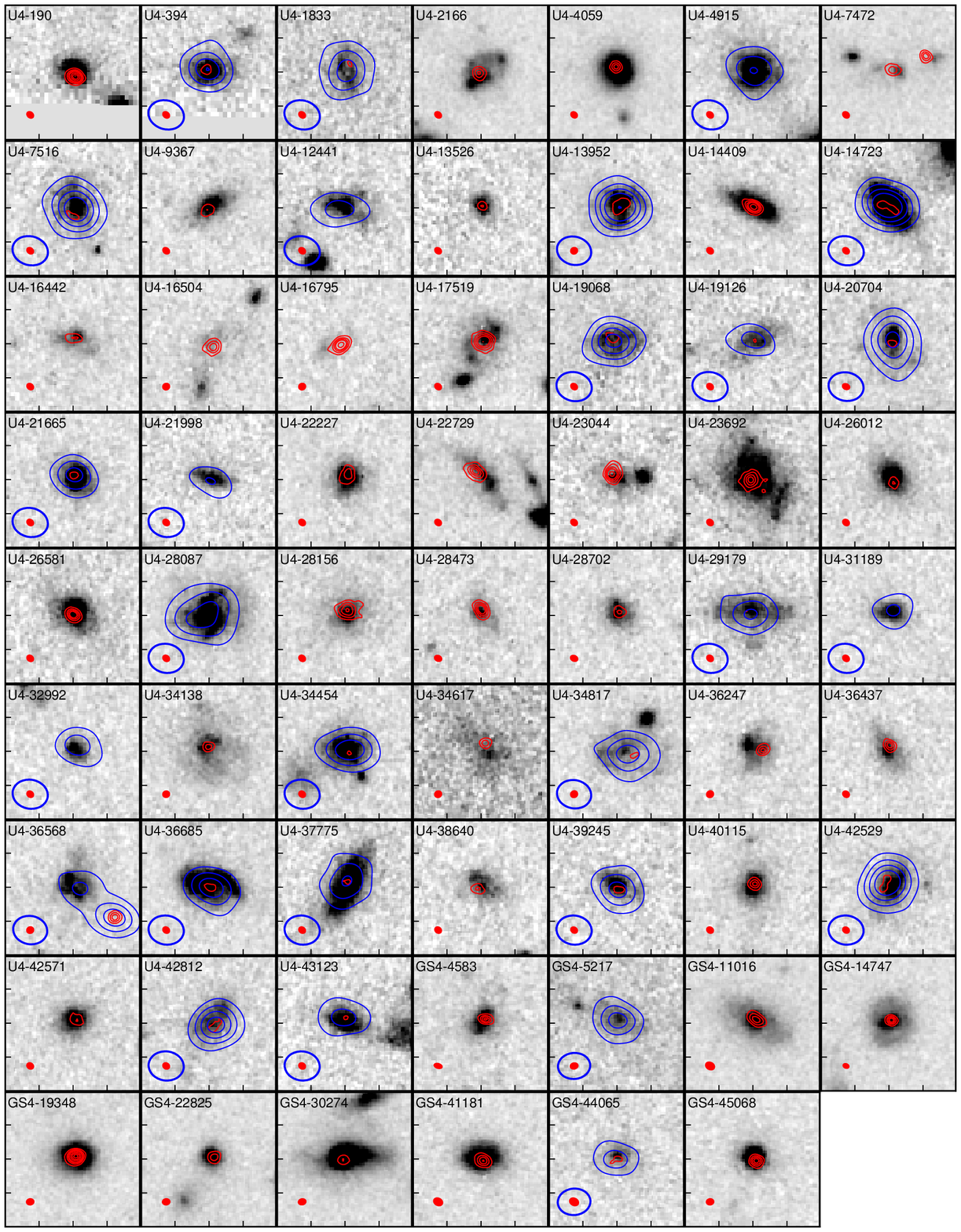}
\end{center}
\caption{
The {\it HST}/F160W-band images (4\arcsec$\times$4\arcsec) for our ALMA sample of 62 massive galaxies.
Red contours display the 870 $\mu$m flux densities in the ALMA high-resolution images.
For galaxies detected at S/N$<10$ in the high-resolution images, we overlay blue contours of the 870 $\mu$m flux densities in the intermediate-resolution images.
Both contours are plotted every 5$\sigma$ to 20$\sigma$ and every 10$\sigma$ from 20$\sigma$.
Red and blue ellipticals in the bottom left corner show the synthesized beams of the high-resolution and the intermediate-resolution images, respectively.
The geometric offset between ALMA and {\it HST} astrometry was corrected in these images (section \ref{sec;vis-fitting}).
}
\label{fig;HSTimage}
\end{figure*}

\begin{figure*}[t]
\begin{center}
\includegraphics[scale=1]{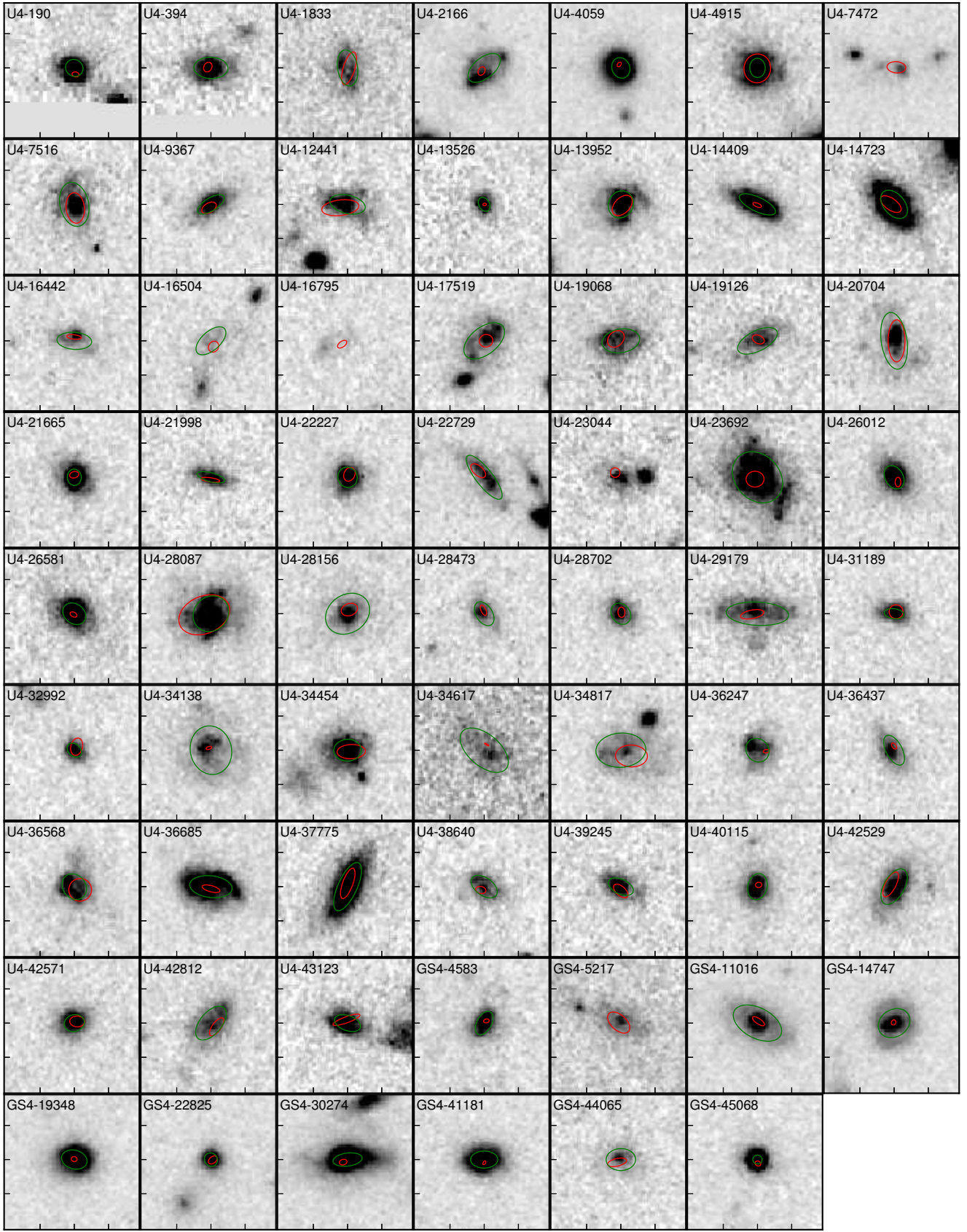}
\end{center}
\caption{
The same as Figure \ref{fig;HSTimage}.
Red and green ellipses are based on the structural parameters (centroid, effective radius, minor-to-major axis ratio and position angle) from the best-fitting S$\acute{\mathrm{e}}$rsic profile in the ALMA/870 $\mu$m and the {\it HST}/F160W-band data, respectively.
}
\label{fig;HSTimage2}
\end{figure*}

We present the catalogs for our ALMA sample (Table \ref{sourcelist} and Table \ref{serendipitous_sourcelist}).
Figure \ref{fig;HSTimage} shows the spatial distributions of the 870 $\mu$m continuum emission on the {\it HST}/F160W-band images for 62 galaxies detected at S/N$>10$ in the ALMA intermediate-resolution images.
As 27 of them are not detected at S/N$>$10 in the high-resolution images, we overlay the contours of the 870 $\mu$m emission in the intermediate-resolution images, that are sensitive to an extended component. 
The 870 $\mu$m emission is smoothly distributed and does not have multiple components like star-forming clumps in disks \citep[e.g.,][]{2016ApJ...833..103H,2019ApJ...882..107R}.
The centrally-concentrated component is the primary star-forming region in massive SFGs as most of star formation is obscured by dust.
Figure \ref{fig;HSTimage2} shows the comparison of the effective radius between ALMA/870 $\mu$m and {\it HST}/F160W-band, demonstrating the result that the dust emission is more compact than the optical emission.

\section*{Appendix B}
\section*{Measurements of structural parameters through visibility fitting}

\begin{figure*}
\begin{center}
\includegraphics[scale=1]{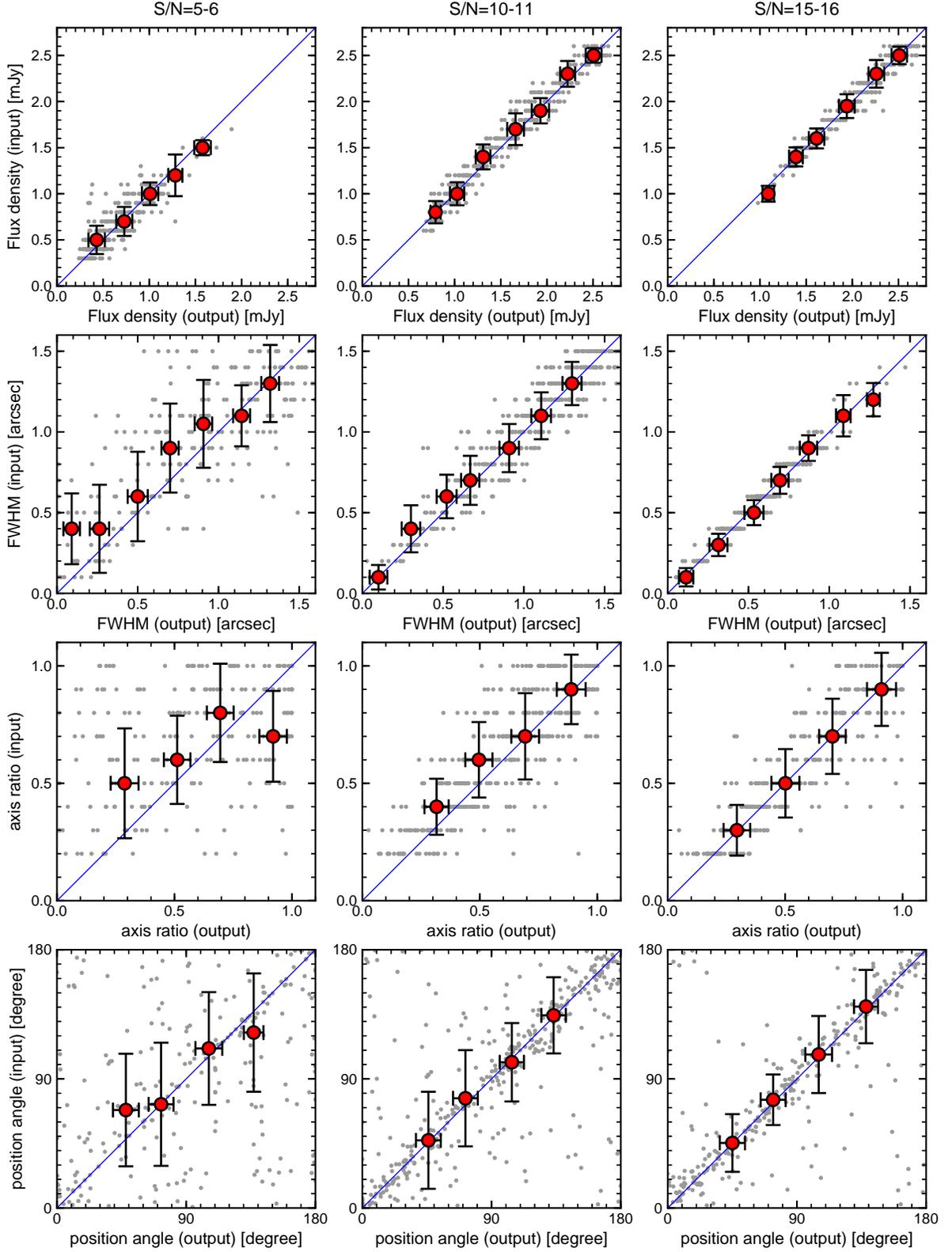}
\end{center}
\caption{
Comparisons of structural parameters (flux density, FWHM, minor-to-major axis ratio and position angle) between measurements by fitting to the simulated visibility data (output) and input models in 6280 mock sources.
Left, middle and right panels show the simulation results for sources detected at $S/N=5-6$, $S/N=10-11$ and $S/N=15-16$ in the intermediate-resolution images, respectively.
Red circles and the error bars indicate the median and the standard deviation in the bins of output values.
}
\label{fig;sim}
\end{figure*}

Using the {\tt CASA Toolkit}, we simulate ALMA observations to evaluate the uncertainties of the size measurements through visibility fitting.
First, we generate elliptical Gaussian models with six free parameters: centroid position, flux density, full width at half maximum (FWHM), minor-to-major axis ratio $q_\mathrm{FIR}$, and position angle PA$_\mathrm{FIR}$.
The position angle is defined as counter-clockwise from North in the images.
Next, we add them as a mock source to the observed visibility data for 2 non-detected objects (U4-17264 and U4-21087).
The flux density varies from 0.3 mJy to 2.6 mJy with a step of  0.1 mJy, the FWHM varies from 0\farcs1 to 1\farcs5 with a step of 0\farcs1 and the axis ratio varies from 0.2 to 1.0 with a step of 0.1, resulting in a total of 6280 mock sources.
The centroid position is random within the range from -3\arcsec to +3\arcsec from the phase center and the position angle is random.
Then, we create intermediate-resolution images in the same way as in Section \ref{sec;ALMAobs} to derive the S/N of the mock sources.
When the mock source is detected at S/N$>$5, we fit elliptical Gaussian models to the visibility data by using the {\tt UVMULTIFIT}.
The structural parameters derived from the best-fit model are compared to the input parameters in three bins: S/N=5--6, S/N=10--11 and S/N=15--16 (Figure \ref{fig;sim}).

The simulations demonstrate the flux measurements work well even at S/N=5 (Figure \ref{fig;sim}).
The uncertainties in the size measurements are typically $\Delta$FWHM=$\pm$0\farcs2 at S/N=5, $\Delta$FWHM=$\pm$0\farcs1 at S/N=10, and $\Delta$FWHM=$\pm$0\farcs05 at S/N=15.
Only when the measured size is as small as FWHM=0\farcs1 at S/N=5, it is underestimated by a factor of 4.
Measuring the axis ratio requires at least S/N=10 where the uncertainties are $\Delta q$=$\pm0.15$.
It seems to be reasonable to adopt circular Gaussian models for measuring the size of the dust emission in 5$<$S/N$<$10 sources.
The important thing here is that there are no systematic errors in the measurement of size over a wide range from FWHM=0\farcs1 to 1\farcs5 in S/N$>$10 conditions.
We therefore use only S/N$>$10 sources for discussion of the size measurements at 870 $\mu$m.

When more than one source is detected in the observed field, we use two- or three-component models for the fitting.
We simply assume circular Gaussian models for the second and the third component to reduce the free parameters.
Even when taking into account the contribution of additional sources, the measurement results only change one percent, except for the two cases where a bright companion is located within 1\farcs5 of the primary targets (U4-7472 and U4-36568).

\begin{figure*}[h]
\begin{center}
\includegraphics[scale=1]{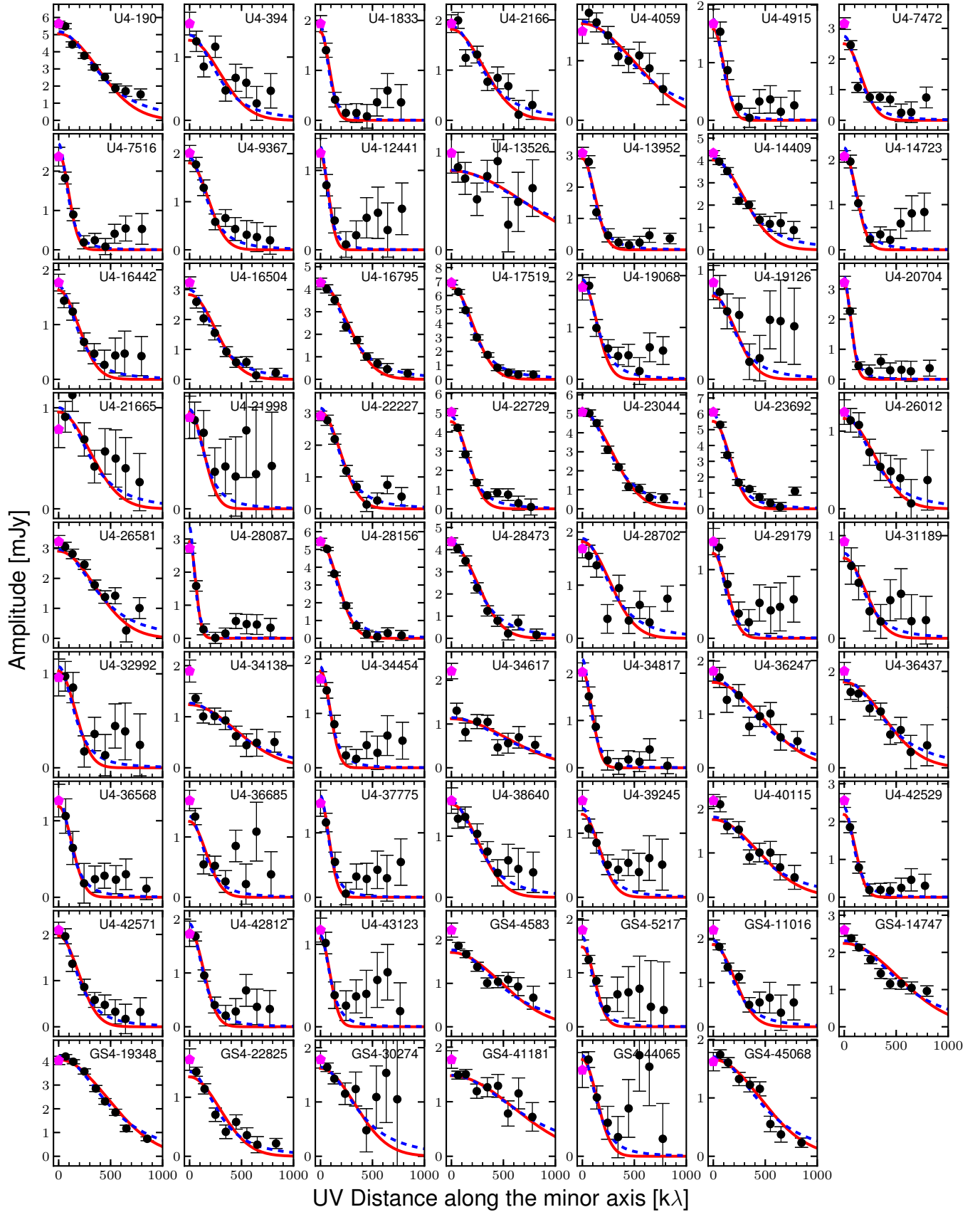}
\end{center}
\caption{
Visibility amplitudes versus $uv$ distances for our sample of 62 massive SFGs detected at $S/N>10$ in the intermediate-resolution images. 
The symbols are the same as Figure \ref{fig;compact_extended}.
}
\label{fig;uvamp}
\end{figure*}

For visualization of the visibility fitting, we compute the $uv$ distances along the minor axis $u'v'$ as

\begin{eqnarray}
u' &=& u\cos \theta - v \sin \theta, \nonumber \\
v' &=& (u\sin \theta + v \cos \theta)\times q_\mathrm{FIR}, \nonumber \\
u'v' &=& \sqrt{u'^2+v'^2},
\end{eqnarray}

\noindent 
where $\theta$=90+PA$_\mathrm{FIR}$.
The minor axis in the visibility plane corresponds to the major axis in the image.
We extract the spatial frequency ($u$,$v$) and the real/imaginary part of individual visibility by using {\tt CASA/plotms} to derive the $u'v'$ distance and the amplitudes.
The amplitudes are computed as the norm of the visibility vectors, averaged in each $u'v'$ bin.
In the right panels of Figure \ref{fig;compact_extended}, we show the amplitudes of the visibilities as a function of the $u'v'$ distance, for both cases of compact and extended sources.
We note that {\tt UVMULTIFIT} fits individual visibility data, not the averaged one.
In Figure \ref{fig;uvamp}, we show the amplitude plots along with the best-fit models for the ALMA sample of 62 galaxies detected at S/N$>$10.
They are used only for visually checking the fitting results.
The amplitudes of extended sources rapidly decline as $u'v'$ distance while those of compact sources slowly change.
As the Fourier transform of the Gaussian function in the image plane is a Gaussian function in the visibility plane, its spatial extent is derived as FWHM$_\mathrm{image}$ [arcsec]=182/FWHM$_\mathrm{visibility}$ [k$\lambda$].
Our ALMA observations demonstrate that visibility data at a very high spatial frequency ($uv>1000$ k$\lambda$) is inefficient for measuring the size of the dust emission for massive SFGs at $z\sim2$ since most of them have FWHM$_\mathrm{image}>$0\farcs15 corresponding to $uv<$600 k$\lambda$.

\section*{Appendix C}
\section*{Calculations of the central stellar mass and the effective radius}

The calculation of the central stellar mass is not straightforward when based on the three-dimensional profile $\rho$.
Following \citep{2009ApJ...697.1290B}, we deproject the two-dimensional S$\acute{\mathrm{e}}$rsic profile using an Abel transform:

\begin{eqnarray}
\rho(x) \propto x^{1/n-1}\int^{\infty}_{1}\frac{\exp(-b_nx^{1/n}t)}{\sqrt{t^{2n}-1}}dt
\end{eqnarray}

\noindent
where $x=r/R_{\mathrm{eff,circ}}$ and $R_{\mathrm{eff,circ}}$ is given by $R_{\mathrm{eff}}\sqrt{q}$. 
The mass of newly formed stars is given by equation \ref{eq;newstar}.
Then, we numerically compute the central mass as 

\begin{eqnarray}
M_\mathrm{1kpc}=\int^{\mathrm{1kpc}}_0 \rho_\mathrm{{\it HST}}(x_\mathrm{{\it HST}})4\pi r^2dr\nonumber \\
+\int^{\mathrm{1kpc}}_0 \rho_\mathrm{ALMA}(x_\mathrm{ALMA})4\pi r^2dr.
\end{eqnarray}

\noindent
where $\rho_\mathrm{{\it HST}}$ and $\rho_\mathrm{{\it ALMA}}$ are the 3D mass profile of already existing and newly formed stars, respectively.
The time evolution of the circularized effective radius can be inferred from the 2D mass profiles ($\Sigma_\mathrm{{\it HST}}$ and $\Sigma_\mathrm{ALMA}$), which are derived from the best-fitting S$\acute{\mathrm{e}}$rsic profile in the {\it HST} and ALMA data, respectively.
We numerically compute the circularized effective radius that satisfies

\begin{eqnarray}
\frac{M_\star +M_{_\star ,\mathrm{new}}}{2}=\int^{R_\mathrm{eff,circ}}_0 \Sigma_\mathrm{{\it HST}}(r)2\pi rdr \nonumber \\
+\int^{R_\mathrm{eff,circ}}_0 \Sigma_\mathrm{ALMA}(r)2\pi rdr.
\end{eqnarray}

\bibliographystyle{apj}

\clearpage

\LongTables
\begin{deluxetable*}{lcccccccccc}
\tablecaption{The source list of 85 massive star-forming galaxies. \label{sourcelist}}
\tablehead{
 ID \tablenotemark{a} & R.A. & Decl. & z \tablenotemark{b} & $\log~M_\star$ & $\log$ SFR  & $R_\mathrm{e,opt}$ \tablenotemark{c} & $q_\mathrm{opt}$ \tablenotemark{c} & S/N \tablenotemark{d} & $S_\mathrm{image}$ \tablenotemark{e} & $\log~M_\mathrm{gas}$ \\
 & (degree) & (degree) & &  ($M_\sun$) & ($M_\sun$yr$^{-1}$)& (kpc)&  & & (mJy) & ($M_\sun$)
}
U4-190 & 34.2654 & -5.2776 &  2.06 & 11.23 & 3.14 & 2.2 & 0.92 & 80.2 & 5.64 $\pm$ 0.14 & 11.48 \\
U4-394 & 34.3965 & -5.2768 &  2.28 & 11.11 & 2.68 & 4.1 & 0.62 & 20.0 & 1.53 $\pm$ 0.18 & 10.92 \\
U4-1620 & 34.3582 & -5.2726 & 2.3185 & 11.27 & 0.27 & 1.3 & 0.90 & 2.4 & $<$0.83  & $<$10.65 \\
U4-1833 & 34.3766 & -5.2713 &  2.51 & 11.28 & 2.72 & 4.3 & 0.55 & 17.7 & 1.94 $\pm$ 0.16 & 11.02 \\
U4-2166 & 34.4264 & -5.2710 &  2.11 & 11.07 & 3.03 & 4.6 & 0.54 & 29.1 & 1.93 $\pm$ 0.18 & 11.01 \\
U4-2394 & 34.5096 & -5.2695 &  2.31 & 11.11 & 2.60 & 2.0 & 0.51 & 6.9 & 0.61 $\pm$ 0.19 & 10.52 \\
U4-4059 & 34.4374 & -5.2642 & 2.323 & 11.09 & 2.57 & 2.5 & 0.85 & 24.2 & 1.49 $\pm$ 0.20 & 10.91 \\
U4-4499 & 34.5142 & -5.2615 & 2.1813 & 11.16 & -0.54 & 0.7 & 0.54 & 2.3 & $<$0.74  & $<$10.60 \\
U4-4701 & 34.2733 & -5.2621 & 2.1065 & 11.49 & -0.22 & 1.8 & 0.62 & 1.9 & $<$0.98  & $<$10.72 \\
U4-4706 & 34.2882 & -5.2626 & 2.5217 & 11.24 & 2.17 & $\cdots$ & $\cdots$ & 5.2 & 0.44 $\pm$ 0.17 & 10.38 \\
U4-4915 & 34.3478 & -5.2612 & 1.907 & 11.13 & 2.26 & 2.4 & 0.80 & 15.1 & 1.68 $\pm$ 0.25 & 10.95 \\
U4-7472 & 34.4654 & -5.2520 & 2.093 & 11.52 & 2.41 & $\cdots$ & $\cdots$ & 34.4 & 3.16 $\pm$ 0.19 & 11.23 \\
U4-7516 & 34.4899 & -5.2524 & 2.080 & 11.30 & 2.26 & 5.4 & 0.64 & 25.6 & 2.37 $\pm$ 0.24 & 11.10 \\
U4-9367 & 34.3316 & -5.2453 & 2.009 & 11.25 & 2.25 & 3.7 & 0.42 & 24.9 & 2.01 $\pm$ 0.16 & 11.03 \\
U4-12441 & 34.5847 & -5.2348 &  1.99 & 11.12 & 2.31 & 4.4 & 0.51 & 12.5 & 1.22 $\pm$ 0.18 & 10.81 \\
U4-13526 & 34.5248 & -5.2308 &  2.49 & 11.21 & 2.32 & 1.7 & 0.72 & 13.6 & 0.99 $\pm$ 0.21 & 10.73 \\
U4-13952 & 34.3226 & -5.2300 & 2.1830 & 11.32 & 2.24 & 3.4 & 0.89 & 30.0 & 3.08 $\pm$ 0.18 & 11.22 \\
U4-14409 & 34.3070 & -5.2281 &  2.17 & 11.21 & 2.38 & 4.8 & 0.36 & 64.8 & 4.33 $\pm$ 0.11 & 11.37 \\
U4-14723 & 34.5298 & -5.2277 &  1.92 & 11.41 & 2.57 & 4.0 & 0.63 & 28.5 & 2.07 $\pm$ 0.13 & 11.04 \\
U4-14996 & 34.2789 & -5.2266 & 2.082 & 11.32 & 1.78 & 2.3 & 0.80 & 6.6 & 0.68 $\pm$ 0.13 & 10.56 \\
U4-16022 & 34.5520 & -5.2223 &  2.40 & 11.34 & 2.00 & 6.5 & 0.19 & 6.0 & 0.46 $\pm$ 0.14 & 10.39 \\
U4-16442 & 34.3367 & -5.2211 & 2.193 & 11.20 & 2.22 & 4.2 & 0.49 & 25.2 & 1.77 $\pm$ 0.15 & 10.98 \\
U4-16504 & 34.4213 & -5.2208 & 2.530 & 11.25 & 2.38 & 4.3 & 0.51 & 35.0 & 3.23 $\pm$ 0.19 & 11.25 \\
U4-16795 & 34.4188 & -5.2196 & 2.530 & 11.18 & 2.62 & $\cdots$ & $\cdots$ & 55.0 & 4.29 $\pm$ 0.20 & 11.37 \\
U4-17264 & 34.4058 & -5.2189 & 2.2986 & 11.14 & -0.56 & 0.9 & 0.95 & 2.9 & $<$0.87  & $<$10.67 \\
U4-17519 & 34.3334 & -5.2183 & 2.385 & 11.60 & 2.89 & 5.6 & 0.57 & 92.0 & 6.92 $\pm$ 0.17 & 11.57 \\
U4-19068 & 34.4233 & -5.2129 &  2.18 & 11.15 & 2.35 & 4.6 & 0.63 & 22.2 & 1.77 $\pm$ 0.24 & 10.98 \\
U4-19126 & 34.5870 & -5.2126 &  2.03 & 11.23 & 2.21 & 5.3 & 0.47 & 12.7 & 0.88 $\pm$ 0.16 & 10.67 \\
U4-20704 & 34.2708 & -5.2079 & 2.1915 & 11.46 & 2.36 & 6.9 & 0.45 & 21.6 & 3.22 $\pm$ 0.21 & 11.24 \\
U4-21087 & 34.2832 & -5.2060 & 2.3168 & 11.11 & -0.59 & 1.0 & 0.94 & 1.8 & $<$0.76  & $<$10.61 \\
U4-21665 & 34.4902 & -5.2040 & 1.964 & 11.03 & 1.83 & 2.0 & 0.85 & 16.9 & 0.79 $\pm$ 0.18 & 10.62 \\
U4-21998 & 34.5178 & -5.2023 &  2.11 & 11.05 & 1.81 & 3.5 & 0.32 & 10.2 & 0.64 $\pm$ 0.15 & 10.53 \\
U4-22227 & 34.4242 & -5.2019 & 2.2896 & 11.27 & 2.36 & 2.5 & 0.94 & 42.7 & 2.92 $\pm$ 0.15 & 11.20 \\
U4-22729 & 34.4693 & -5.2010 & 2.579 & 11.34 & 2.75 & 6.4 & 0.31 & 63.9 & 5.03 $\pm$ 0.18 & 11.44 \\
U4-23044 & 34.3857 & -5.1990 & 2.421 & 11.61 & 3.08 & $\cdots$ & $\cdots$ & 76.4 & 5.09 $\pm$ 0.20 & 11.44 \\
U4-23692 & 34.3632 & -5.1994 & 2.031 & 11.64 & 2.91 & 6.9 & 0.79 & 72.1 & 6.11 $\pm$ 0.18 & 11.51 \\
U4-26012 & 34.2653 & -5.1895 & 2.3208 & 11.06 & 2.41 & 3.0 & 0.72 & 17.8 & 1.25 $\pm$ 0.16 & 10.83 \\
U4-26581 & 34.5491 & -5.1877 &  2.22 & 11.24 & 2.50 & 3.0 & 0.79 & 45.1 & 3.23 $\pm$ 0.20 & 11.24 \\
U4-28087 & 34.3588 & -5.1828 & 2.113 & 11.34 & 2.30 & 4.6 & 0.89 & 17.1 & 2.72 $\pm$ 0.15 & 11.16 \\
U4-28156 & 34.4685 & -5.1824 & 1.995 & 11.36 & 2.51 & 5.7 & 0.81 & 71.1 & 5.44 $\pm$ 0.19 & 11.46 \\
U4-28473 & 34.4224 & -5.1810 & 2.5247 & 11.35 & 2.59 & 3.1 & 0.62 & 54.7 & 4.36 $\pm$ 0.17 & 11.38 \\
U4-28702 & 34.4775 & -5.1800 & 2.190 & 11.03 & 2.10 & 2.7 & 0.81 & 22.9 & 1.69 $\pm$ 0.17 & 10.96 \\
U4-29179 & 34.5888 & -5.1789 & 2.006 & 11.39 & 2.10 & 7.7 & 0.37 & 15.9 & 1.41 $\pm$ 0.19 & 10.87 \\
U4-30882 & 34.2294 & -5.1730 & 2.1057 & 11.01 & 0.01 & 0.9 & 0.66 & 1.1 & $<$0.91  & $<$10.69 \\
U4-31189 & 34.4158 & -5.1719 & 2.294 & 11.16 & 2.40 & 2.4 & 0.51 & 10.8 & 0.94 $\pm$ 0.19 & 10.71 \\
U4-32147 & 34.3996 & -5.1693 & 2.5294 & 11.15 & 1.90 & 4.5 & 0.52 & 8.5 & 0.84 $\pm$ 0.16 & 10.66 \\
U4-32351 & 34.3061 & -5.1682 & 2.1800 & 11.04 & 2.17 & 3.4 & 0.70 & 9.3 & 0.84 $\pm$ 0.18 & 10.65 \\
U4-32992 & 34.4727 & -5.1657 & 2.182 & 11.01 & 2.21 & 2.0 & 0.74 & 13.1 & 0.96 $\pm$ 0.20 & 10.71 \\
U4-34138 & 34.4086 & -5.1631 & 2.5186 & 11.05 & 2.23 & 5.8 & 0.84 & 18.3 & 1.90 $\pm$ 0.22 & 11.01 \\
U4-34454 & 34.2867 & -5.1615 & 2.027 & 11.18 & 2.36 & 3.4 & 0.79 & 19.4 & 1.63 $\pm$ 0.13 & 10.94 \\
U4-34617 & 34.4785 & -5.1606 & 2.530 & 11.04 & 2.42 & 6.8 & 0.55 & 17.6 & 2.21 $\pm$ 0.18 & 11.08 \\
U4-34817 & 34.2718 & -5.1601 & 2.190 & 11.21 & 2.37 & 6.0 & 0.69 & 18.0 & 2.02 $\pm$ 0.19 & 11.04 \\
U4-36247 & 34.2817 & -5.1548 & 2.1790 & 11.04 & 2.41 & 3.0 & 0.89 & 23.3 & 1.80 $\pm$ 0.19 & 10.99 \\
U4-36437 & 34.4852 & -5.1542 & 2.083 & 11.13 & 2.47 & 4.0 & 0.49 & 27.2 & 2.01 $\pm$ 0.18 & 11.03 \\
U4-36568 & 34.2938 & -5.1545 & 2.1770 & 11.05 & 2.40 & 3.3 & 0.84 & 10.9 & 1.26 $\pm$ 0.21 & 10.83 \\
U4-36685 & 34.2893 & -5.1538 &  1.95 & 11.38 & 2.48 & 5.2 & 0.51 & 19.0 & 1.60 $\pm$ 0.17 & 10.93 \\
U4-37775 & 34.5325 & -5.1503 &  1.92 & 11.49 & 2.43 & 6.4 & 0.36 & 15.5 & 1.46 $\pm$ 0.20 & 10.89 \\
U4-38040 & 34.4687 & -5.1481 &  2.49 & 11.02 & 2.08 & 3.6 & 0.49 & 7.6 & 0.43 $\pm$ 0.17 & 10.37 \\
U4-38640 & 34.5595 & -5.1460 &  2.42 & 11.25 & 2.39 & 3.5 & 0.59 & 21.5 & 1.58 $\pm$ 0.21 & 10.93 \\
U4-39126 & 34.5498 & -5.1444 &  2.02 & 11.13 & 2.08 & 6.8 & 0.31 & 6.3 & 0.54 $\pm$ 0.19 & 10.46 \\
U4-39245 & 34.5539 & -5.1438 &  2.12 & 11.14 & 2.42 & 3.2 & 0.55 & 17.3 & 1.51 $\pm$ 0.16 & 10.91 \\
U4-39537 & 34.4820 & -5.1437 & 2.0234 & 11.24 & 2.09 & 2.2 & 0.82 & 3.7 & $<$0.69  & $<$10.56 \\
U4-40115 & 34.4408 & -5.1411 &  2.02 & 11.08 & 2.47 & 3.2 & 0.72 & 28.6 & 2.18 $\pm$ 0.14 & 11.06 \\
U4-41248 & 34.3150 & -5.1356 & 1.9210 & 11.01 & 0.82 & 1.3 & 0.84 & 2.5 & $<$0.91  & $<$10.68 \\
U4-42529 & 34.4454 & -5.1305 &  2.39 & 11.53 & 2.58 & 4.7 & 0.52 & 26.4 & 2.56 $\pm$ 0.18 & 11.14 \\
U4-42571 & 34.4331 & -5.1309 &  2.41 & 11.38 & 2.52 & 2.4 & 0.82 & 26.9 & 2.10 $\pm$ 0.17 & 11.06 \\
U4-42812 & 34.2675 & -5.1296 &  2.12 & 11.16 & 2.41 & 4.9 & 0.54 & 23.0 & 1.73 $\pm$ 0.24 & 10.97 \\
U4-43123 & 34.5208 & -5.1289 &  2.03 & 11.04 & 2.31 & 3.3 & 0.55 & 12.1 & 1.17 $\pm$ 0.14 & 10.80 \\
U4-43667 & 34.2584 & -5.1266 &  2.04 & 11.15 & 1.90 & 5.8 & 0.68 & 6.4 & 0.92 $\pm$ 0.15 & 10.69 \\
GS4-1397 & 53.1040 & -27.9226 & 2.5552 & 11.03 & 0.58 & 3.5 & 0.48 & 3.5 & $<$0.84  & $<$10.66 \\
GS4-1725 & 53.2049 & -27.9179 & 2.206 & 11.03 & 2.31 & 2.8 & 0.66 & 9.0 & 0.78 $\pm$ 0.28 & 10.62 \\
GS4-2467 & 53.2112 & -27.9072 & 2.0404 & 11.08 & 1.48 & $\cdots$ & $\cdots$ & 1.1 & $<$0.86  & $<$10.66 \\
GS4-4583 & 53.1635 & -27.8905 & 2.281 & 11.11 & 2.64 & 3.0 & 0.58 & 32.3 & 2.23 $\pm$ 0.16 & 11.08 \\
GS4-5217 & 53.1131 & -27.8866 & 2.426 & 11.02 & 1.92 & $\cdots$ & $\cdots$ & 17.7 & 1.79 $\pm$ 0.18 & 10.99 \\
GS4-11016 & 53.0773 & -27.8596 & 2.0369 & 11.24 & 1.98 & 6.4 & 0.58 & 30.0 & 2.19 $\pm$ 0.22 & 11.07 \\
GS4-14747 & 53.0717 & -27.8436 & 1.9560 & 11.09 & 2.33 & 4.1 & 0.81 & 42.5 & 2.60 $\pm$ 0.17 & 11.14 \\
GS4-19348 & 53.1488 & -27.8211 & 2.5820 & 11.26 & 2.91 & 3.0 & 0.72 & 68.7 & 4.04 $\pm$ 0.17 & 11.34 \\
GS4-22825 & 53.0940 & -27.8041 & 2.367 & 11.15 & 2.03 & 1.7 & 0.91 & 22.1 & 1.64 $\pm$ 0.18 & 10.95 \\
GS4-30274 & 53.1311 & -27.7731 & 2.2250 & 11.20 & 2.53 & 3.5 & 0.44 & 27.0 & 1.78 $\pm$ 0.16 & 10.98 \\
GS4-40185 & 53.0136 & -27.7201 & 2.076 & 11.14 & 1.49 & 6.2 & 0.74 & 5.5 & 0.46 $\pm$ 0.15 & 10.39 \\
GS4-41021 & 53.1874 & -27.7192 & 2.3135 & 11.02 & 2.10 & 0.5 & 0.55 & -0.1 & $<$0.73  & $<$10.60 \\
GS4-41181 & 53.1070 & -27.7182 & 2.3001 & 11.28 & 2.10 & 3.3 & 0.63 & 27.7 & 1.77 $\pm$ 0.16 & 10.98 \\
GS4-44065 & 53.0026 & -27.7044 &  2.59 & 11.09 & 2.29 & 3.4 & 0.75 & 12.4 & 0.73 $\pm$ 0.15 & 10.60 \\
GS4-45068 & 53.1376 & -27.7001 & 2.4480 & 11.01 & 2.56 & 1.1 & 0.94 & 28.7 & 1.62 $\pm$ 0.16 & 10.94 \\
GS4-45475 & 53.1737 & -27.6981 &  2.03 & 11.09 & 1.89 & 8.0 & 0.58 & 6.4 & 0.85 $\pm$ 0.17 & 10.65 
  \enddata
\tablenotetext{a}{Unique identifier in the 3D-HST v4 catalog \citep{2014ApJS..214...24S, 2016ApJS..225...27M}. U4 and GS4 mean objects in the UDS and the GOODS-S field, respectively.}
\tablenotetext{b}{Numbers with 3, 4 and 5 digits means photometric, grism or narrow-band, and spectroscopic redshift, respectively (Section \ref{sec;selection}).}
\tablenotetext{c}{Effective radius along the semi-major axis and the minor-to-major axis ratio from the best-fitting S$\acute{\mathrm{e}}$rsic profile in the {\it HST}/F160W \citep{2012ApJS..203...24V,2014ApJ...788...28V,2014ApJ...788...11L}.}
\tablenotetext{d}{Signal-to-noise ratio in the ALMA intermediate-resolution images.}
\tablenotetext{e}{870 $\mu$m flux densities measured in the ALMA low-resolution images.}
\end{deluxetable*}

\clearpage

\LongTables
\begin{deluxetable*}{lcccccccccc}
\tablecaption{The galaxy properties of 19 serendipitously detected sources \label{serendipitous_sourcelist}}
\tablehead{
 ID & R.A. & Decl. & z \tablenotemark{a} & $\log~M_\star$ & $\log$ SFR  & S/N  & $S_\mathrm{image}$  & Primary source ID\tablenotemark{b} & Separation\tablenotemark{c} & $|\Delta z|$\tablenotemark{d} \\
 & (degree) & (degree) & &  ($M_\sun$) & ($M_\sun$yr$^{-1}$)& & (mJy) & & (arcsec) & 
}
U4-570 & 34.3951 & -5.2759 &  2.27 & 10.87 & 2.30 & 12.6 & 1.04 $\pm$ 0.24  & U4-394 & 6.1 & 0.01\\
U4-4018 & 34.3493 & -5.2634 &  2.98 & 11.37 & 2.94 & 8.9 & 1.75 $\pm$ 0.61  & U4-4915 & 9.6 & 1.07\\
U4-4303 & 34.2879 & -5.2636 & 2.454 & 10.80 & 1.77 & 27.2 & 2.12 $\pm$ 0.20  & U4-4706 & 3.8 & 0.07\\
U4-4534 & 34.4388 & -5.2636 & 0.780 & 10.59 & 1.66 & 5.8 & 0.86 $\pm$ 0.27  & U4-4059 & 5.5 & 1.54\\
U4-7419 & 34.4650 & -5.2519 &  2.97 & 11.24 & 0.80 & 29.6 & 2.36 $\pm$ 0.19  & U4-7472 & 1.3 & 0.88\\
U4-9997 & 34.3293 & -5.2424 &  2.77 & 10.14 & 2.11 & 7.3 & 4.15 $\pm$ 0.93  & U4-9367 & 13.3 & 0.76\\
U4-14449 & 34.2775 & -5.2281 &  1.87 & 10.80 & 2.24 & 12.4 & 1.58 $\pm$ 0.21  & U4-14996 & 7.3 & 0.21\\
U4-17813 & 34.5878 & -5.2161 &  4.67 & 11.69 & 2.04 & 8.4 & 1.85 $\pm$ 0.87  & U4-19126 & 13.1 & 2.64\\
U4-35673 & 34.2722 & -5.1571 & 2.190 & 10.96 & 2.79 & 10.0 & 3.21 $\pm$ 0.58  & U4-34817 & 10.7 & 0.00\\
U4-36568b \tablenotemark{e} & 34.2935 & -5.1548 & $\cdots$ & $\cdots$ &  $\cdots$ & 18.7 & 1.51 $\pm$ 0.21  & U4-36568 & 1.5 & $\cdots$ \\
U4-39404 & 34.5505 & -5.1433 &  1.63 & 10.47 & 1.63 & 5.5 & 0.46 $\pm$ 0.22  & U4-39126 & 4.6 & 0.40\\
GS4-10771 & 53.0785 & -27.8599 & 3.6600 & 11.20 & 2.80 & 8.7 & 0.70 $\pm$ 0.25   & GS4-11016 & 3.8 & 1.62 \\
GS4-11466 & 53.0779 & -27.8582 & 0.6465 & 10.62 & 1.37 & 6.4 & 0.72 $\pm$ 0.29   & GS4-11016 & 5.4 & 1.39 \\
GS4-14173 & 53.0702 & -27.8455 &  3.70 & 10.34 & 1.98 & 18.7 & 2.12 $\pm$ 0.31  & GS4-14747 & 8.3 & 1.74\\
GS4-23075 & 53.0923 & -27.8032 &  2.72 & 11.02 & 2.47 & 25.9 & 2.87 $\pm$ 0.27  & GS4-22825 & 6.6 & 0.35\\
GS4-23372 & 53.0928 & -27.8012 &  2.93 & 10.65 & 2.14 & 18.6 & 3.86 $\pm$ 0.58  & GS4-22825 & 11.2 & 0.56\\
GS4-41332 & 53.1100 & -27.7168 &  3.10 & 10.20 & 2.27 & 10.8 & 1.66 $\pm$ 0.50  & GS4-41181 & 10.8 & 0.80\\
GS4-44920 & 53.1388 & -27.7005 & 2.4495 & 10.49 & 1.83 & 5.4 & 0.39 $\pm$ 0.18   & GS4-45068 & 4.0 & 0.00 \\
GS4-46181 & 53.1753 & -27.6948 &  4.79 & 11.92 & 3.64 & 15.1 & 6.78 $\pm$ 0.92  & GS4-45475 & 13.0 & 2.77
  \enddata
\tablenotetext{a}{Redshifts with 5 digits are based on spectroscopic observations \citep{2004ApJS..155..271S,2009A&A...494..443P,2013A&A...549A..63K}.}
\tablenotetext{b}{3D-HST ID of the primary target in ALMA observations.}
\tablenotetext{c}{Projected separation between the serendipitously detected source and the primary target.}
\tablenotetext{d}{Redshift difference between the serendipitously detected source and the primary target.}
\tablenotetext{e}{U4-36568b is not detected in the HST image and is visible only in the ALMA 870 $\mu$m image.}
\end{deluxetable*}

\clearpage

\LongTables
\begin{deluxetable*}{lcccccc}
\tablecaption{The 870 $\mu$m properties for 74 massive star-forming galaxies. \label{visi-fit}}
\tablehead{
 ID & $S_\mathrm{visibility}$ (n=0.5) & $R_\mathrm{e,FIR}$ (n=0.5) & $q_\mathrm{FIR}$ (n=0.5) & PA$_\mathrm{FIR}$ (n=0.5) & $S_\mathrm{visibility}$ (n=1) & $R_\mathrm{e,FIR}$ (n=1)\\
 & (mJy) &  (kpc) & & & (mJy) &  (kpc)
}
U4-190 & 5.05 $\pm$ 0.09 & 0.9 $\pm$ 0.1 & 0.71 $\pm$ 0.07 & 79 $\pm$ 7 & 5.18 $\pm$ 0.10 & 0.8 $\pm$ 0.0 \\
U4-394 & 1.27 $\pm$ 0.10 & 1.1 $\pm$ 0.2 & 0.78 $\pm$ 0.29 & 152 $\pm$ 42 & 1.35 $\pm$ 0.11 & 1.2 $\pm$ 0.2 \\
U4-1833 & 1.77 $\pm$ 0.16 & 3.8 $\pm$ 0.5 & 0.34 $\pm$ 0.10 & 163 $\pm$ 6 & 1.89 $\pm$ 0.20 & 4.0 $\pm$ 0.6 \\
U4-2166 & 1.81 $\pm$ 0.10 & 1.1 $\pm$ 0.2 & 0.73 $\pm$ 0.20 & 152 $\pm$ 22 & 1.88 $\pm$ 0.11 & 1.1 $\pm$ 0.1 \\
U4-2394 & 0.68 $\pm$ 0.18 & 3.0 $\pm$ 0.9 & $\cdots$ & $\cdots$ & 0.69 $\pm$ 0.22 & 2.9 $\pm$ 1.2 \\
U4-4059 & 1.62 $\pm$ 0.08 & 0.6 $\pm$ 0.1 & 0.67 $\pm$ 0.29 & 145 $\pm$ 27 & 1.65 $\pm$ 0.09 & 0.6 $\pm$ 0.1 \\
U4-4706 & 0.42 $\pm$ 0.15 & 2.3 $\pm$ 1.1 & $\cdots$ & $\cdots$ & 0.27 $\pm$ 0.08 & 0.3 $\pm$ 0.5 \\
U4-4915 & 1.58 $\pm$ 0.19 & 3.3 $\pm$ 0.5 & 0.92 $\pm$ 0.25 & 171 $\pm$ 88 & 1.75 $\pm$ 0.24 & 3.6 $\pm$ 0.6 \\
U4-7472 & 2.50 $\pm$ 0.12 & 2.1 $\pm$ 0.2 & 0.60 $\pm$ 0.11 & 85 $\pm$ 9 & 2.74 $\pm$ 0.15 & 2.3 $\pm$ 0.2 \\
U4-7516 & 2.46 $\pm$ 0.17 & 3.4 $\pm$ 0.3 & 0.60 $\pm$ 0.11 & 6 $\pm$ 9 & 2.69 $\pm$ 0.22 & 3.7 $\pm$ 0.4 \\
U4-9367 & 1.81 $\pm$ 0.12 & 1.9 $\pm$ 0.3 & 0.60 $\pm$ 0.15 & 121 $\pm$ 14 & 1.91 $\pm$ 0.14 & 1.9 $\pm$ 0.3 \\
U4-12441 & 1.19 $\pm$ 0.18 & 4.3 $\pm$ 0.9 & 0.42 $\pm$ 0.16 & 97 $\pm$ 10 & 1.29 $\pm$ 0.22 & 4.6 $\pm$ 1.0 \\
U4-13526 & 0.81 $\pm$ 0.08 & 0.4 $\pm$ 0.2 & 0.68 $\pm$ 0.85 & 98 $\pm$ 80 & 0.81 $\pm$ 0.08 & 0.4 $\pm$ 0.2 \\
U4-13952 & 2.91 $\pm$ 0.17 & 2.7 $\pm$ 0.3 & 0.64 $\pm$ 0.11 & 136 $\pm$ 10 & 3.16 $\pm$ 0.22 & 2.9 $\pm$ 0.3 \\
U4-14409 & 4.02 $\pm$ 0.09 & 1.1 $\pm$ 0.1 & 0.40 $\pm$ 0.06 & 68 $\pm$ 3 & 4.12 $\pm$ 0.10 & 1.1 $\pm$ 0.1 \\
U4-14723 & 2.14 $\pm$ 0.14 & 2.9 $\pm$ 0.3 & 0.48 $\pm$ 0.10 & 55 $\pm$ 7 & 2.26 $\pm$ 0.17 & 2.9 $\pm$ 0.3 \\
U4-14996 & 0.42 $\pm$ 0.09 & 0.7 $\pm$ 0.3 & $\cdots$ & $\cdots$ & 0.44 $\pm$ 0.10 & 0.7 $\pm$ 0.4 \\
U4-16022 & 0.32 $\pm$ 0.08 & 0.3 $\pm$ 0.4 & $\cdots$ & $\cdots$ & 0.32 $\pm$ 0.08 & 0.3 $\pm$ 0.4 \\
U4-16442 & 1.62 $\pm$ 0.10 & 1.7 $\pm$ 0.2 & 0.29 $\pm$ 0.12 & 85 $\pm$ 6 & 1.70 $\pm$ 0.11 & 1.8 $\pm$ 0.3 \\
U4-16504 & 2.83 $\pm$ 0.12 & 1.3 $\pm$ 0.1 & 0.85 $\pm$ 0.14 & 146 $\pm$ 29 & 2.99 $\pm$ 0.14 & 1.3 $\pm$ 0.1 \\
U4-16795 & 4.22 $\pm$ 0.11 & 1.2 $\pm$ 0.1 & 0.53 $\pm$ 0.06 & 126 $\pm$ 5 & 4.39 $\pm$ 0.12 & 1.2 $\pm$ 0.1 \\
U4-17519 & 6.61 $\pm$ 0.12 & 1.6 $\pm$ 0.1 & 0.92 $\pm$ 0.06 & 108 $\pm$ 22 & 6.98 $\pm$ 0.14 & 1.6 $\pm$ 0.1 \\
U4-19068 & 1.87 $\pm$ 0.14 & 2.4 $\pm$ 0.3 & 0.76 $\pm$ 0.18 & 135 $\pm$ 24 & 1.93 $\pm$ 0.17 & 2.3 $\pm$ 0.3 \\
U4-19126 & 0.76 $\pm$ 0.10 & 1.6 $\pm$ 0.5 & 0.64 $\pm$ 0.38 & 65 $\pm$ 35 & 0.79 $\pm$ 0.12 & 1.6 $\pm$ 0.5 \\
U4-20704 & 3.15 $\pm$ 0.24 & 4.9 $\pm$ 0.5 & 0.40 $\pm$ 0.09 & 180 $\pm$ 5 & 3.40 $\pm$ 0.29 & 5.1 $\pm$ 0.6 \\
U4-21665 & 0.96 $\pm$ 0.09 & 1.1 $\pm$ 0.3 & 0.68 $\pm$ 0.34 & 106 $\pm$ 36 & 1.00 $\pm$ 0.11 & 1.1 $\pm$ 0.3 \\
U4-21998 & 0.67 $\pm$ 0.11 & 2.4 $\pm$ 0.8 & 0.23 $\pm$ 0.24 & 82 $\pm$ 11 & 0.69 $\pm$ 0.12 & 2.3 $\pm$ 0.8 \\
U4-22227 & 3.04 $\pm$ 0.12 & 1.7 $\pm$ 0.1 & 0.76 $\pm$ 0.11 & 156 $\pm$ 14 & 3.16 $\pm$ 0.14 & 1.7 $\pm$ 0.1 \\
U4-22729 & 4.53 $\pm$ 0.12 & 2.0 $\pm$ 0.1 & 0.53 $\pm$ 0.05 & 48 $\pm$ 4 & 4.82 $\pm$ 0.14 & 2.1 $\pm$ 0.1 \\
U4-23044 & 5.04 $\pm$ 0.09 & 1.1 $\pm$ 0.1 & 0.99 $\pm$ 0.08 & 0 $\pm$ 2 & 5.22 $\pm$ 0.12 & 1.1 $\pm$ 0.1 \\
U4-23692 & 5.52 $\pm$ 0.14 & 2.1 $\pm$ 0.1 & 0.88 $\pm$ 0.07 & 92 $\pm$ 17 & 5.97 $\pm$ 0.17 & 2.2 $\pm$ 0.1 \\
U4-26012 & 1.16 $\pm$ 0.10 & 1.1 $\pm$ 0.2 & 0.58 $\pm$ 0.26 & 180 $\pm$ 20 & 1.21 $\pm$ 0.11 & 1.2 $\pm$ 0.2 \\
U4-26581 & 2.91 $\pm$ 0.09 & 0.9 $\pm$ 0.1 & 0.62 $\pm$ 0.11 & 63 $\pm$ 10 & 2.96 $\pm$ 0.10 & 0.9 $\pm$ 0.1 \\
U4-28087 & 2.92 $\pm$ 0.27 & 5.6 $\pm$ 0.6 & 0.70 $\pm$ 0.11 & 113 $\pm$ 12 & 3.35 $\pm$ 0.37 & 6.4 $\pm$ 0.7 \\
U4-28156 & 5.33 $\pm$ 0.12 & 2.0 $\pm$ 0.1 & 0.72 $\pm$ 0.06 & 109 $\pm$ 7 & 5.61 $\pm$ 0.15 & 2.0 $\pm$ 0.1 \\
U4-28473 & 4.22 $\pm$ 0.12 & 1.3 $\pm$ 0.1 & 0.44 $\pm$ 0.06 & 24 $\pm$ 4 & 4.35 $\pm$ 0.13 & 1.3 $\pm$ 0.1 \\
U4-28702 & 1.82 $\pm$ 0.12 & 1.3 $\pm$ 0.2 & 0.67 $\pm$ 0.23 & 3 $\pm$ 23 & 1.88 $\pm$ 0.14 & 1.2 $\pm$ 0.2 \\
U4-29179 & 1.24 $\pm$ 0.12 & 2.6 $\pm$ 0.5 & 0.36 $\pm$ 0.14 & 101 $\pm$ 9 & 1.35 $\pm$ 0.15 & 2.9 $\pm$ 0.6 \\
U4-31189 & 0.78 $\pm$ 0.12 & 1.6 $\pm$ 0.4 & $\cdots$ & $\cdots$ & 0.83 $\pm$ 0.15 & 1.7 $\pm$ 0.5 \\
U4-32147 & 0.82 $\pm$ 0.18 & 2.8 $\pm$ 0.7 & $\cdots$ & $\cdots$ & 0.87 $\pm$ 0.22 & 2.7 $\pm$ 0.9 \\
U4-32351 & 0.67 $\pm$ 0.10 & 0.7 $\pm$ 0.2 & $\cdots$ & $\cdots$ & 0.70 $\pm$ 0.11 & 0.7 $\pm$ 0.2 \\
U4-32992 & 1.03 $\pm$ 0.13 & 2.1 $\pm$ 0.5 & 0.67 $\pm$ 0.30 & 166 $\pm$ 29 & 1.08 $\pm$ 0.15 & 2.1 $\pm$ 0.5 \\
U4-34138 & 1.24 $\pm$ 0.10 & 0.7 $\pm$ 0.2 & 0.47 $\pm$ 0.25 & 107 $\pm$ 16 & 1.27 $\pm$ 0.10 & 0.7 $\pm$ 0.1 \\
U4-34454 & 1.78 $\pm$ 0.16 & 3.5 $\pm$ 0.5 & 0.50 $\pm$ 0.13 & 92 $\pm$ 9 & 1.86 $\pm$ 0.19 & 3.5 $\pm$ 0.5 \\
U4-34617 & 1.12 $\pm$ 0.09 & 0.6 $\pm$ 0.1 & 0.27 $\pm$ 0.33 & 60 $\pm$ 12 & 1.15 $\pm$ 0.09 & 0.6 $\pm$ 0.1 \\
U4-34817 & 2.04 $\pm$ 0.21 & 3.5 $\pm$ 0.6 & 0.68 $\pm$ 0.17 & 88 $\pm$ 18 & 2.29 $\pm$ 0.28 & 3.9 $\pm$ 0.6 \\
U4-36247 & 1.59 $\pm$ 0.10 & 0.7 $\pm$ 0.1 & 0.55 $\pm$ 0.20 & 98 $\pm$ 15 & 1.66 $\pm$ 0.10 & 0.7 $\pm$ 0.1 \\
U4-36437 & 1.78 $\pm$ 0.09 & 0.9 $\pm$ 0.1 & 0.59 $\pm$ 0.17 & 36 $\pm$ 16 & 1.82 $\pm$ 0.10 & 0.8 $\pm$ 0.1 \\
U4-36568 & 1.19 $\pm$ 0.20 & 2.6 $\pm$ 0.5 & $\cdots$ & $\cdots$ & 1.30 $\pm$ 0.25 & 2.8 $\pm$ 0.7 \\
U4-36685 & 1.26 $\pm$ 0.11 & 2.1 $\pm$ 0.4 & 0.33 $\pm$ 0.15 & 75 $\pm$ 8 & 1.34 $\pm$ 0.13 & 2.3 $\pm$ 0.4 \\
U4-37775 & 1.50 $\pm$ 0.16 & 3.9 $\pm$ 0.6 & 0.32 $\pm$ 0.12 & 160 $\pm$ 7 & 1.57 $\pm$ 0.19 & 3.9 $\pm$ 0.7 \\
U4-38040 & 0.51 $\pm$ 0.11 & 1.3 $\pm$ 0.5 & $\cdots$ & $\cdots$ & 0.57 $\pm$ 0.14 & 1.5 $\pm$ 0.7 \\
U4-38640 & 1.52 $\pm$ 0.10 & 1.2 $\pm$ 0.2 & 0.68 $\pm$ 0.21 & 78 $\pm$ 20 & 1.57 $\pm$ 0.11 & 1.2 $\pm$ 0.2 \\
U4-39126 & 0.46 $\pm$ 0.12 & 1.4 $\pm$ 0.6 & $\cdots$ & $\cdots$ & 0.49 $\pm$ 0.14 & 1.6 $\pm$ 0.8 \\
U4-39245 & 1.29 $\pm$ 0.12 & 2.0 $\pm$ 0.4 & 0.53 $\pm$ 0.18 & 48 $\pm$ 14 & 1.39 $\pm$ 0.14 & 2.1 $\pm$ 0.4 \\
U4-40115 & 1.75 $\pm$ 0.09 & 0.8 $\pm$ 0.1 & 0.84 $\pm$ 0.25 & 99 $\pm$ 45 & 1.81 $\pm$ 0.10 & 0.8 $\pm$ 0.1 \\
U4-42529 & 2.19 $\pm$ 0.15 & 3.1 $\pm$ 0.3 & 0.39 $\pm$ 0.09 & 153 $\pm$ 6 & 2.37 $\pm$ 0.18 & 3.3 $\pm$ 0.4 \\
U4-42571 & 1.97 $\pm$ 0.12 & 1.7 $\pm$ 0.2 & 0.73 $\pm$ 0.17 & 87 $\pm$ 19 & 2.10 $\pm$ 0.14 & 1.8 $\pm$ 0.2 \\
U4-42812 & 1.80 $\pm$ 0.13 & 2.5 $\pm$ 0.3 & 0.41 $\pm$ 0.11 & 144 $\pm$ 7 & 1.91 $\pm$ 0.16 & 2.6 $\pm$ 0.4 \\
U4-43123 & 1.07 $\pm$ 0.14 & 3.6 $\pm$ 0.8 & 0.21 $\pm$ 0.12 & 110 $\pm$ 6 & 1.11 $\pm$ 0.16 & 3.5 $\pm$ 0.8 \\
U4-43667 & 0.48 $\pm$ 0.13 & 1.8 $\pm$ 0.7 & $\cdots$ & $\cdots$ & 0.51 $\pm$ 0.16 & 1.9 $\pm$ 1.0 \\
GS4-1725 & 1.00 $\pm$ 0.19 & 1.7 $\pm$ 0.4 & $\cdots$ & $\cdots$ & 1.04 $\pm$ 0.23 & 1.7 $\pm$ 0.5 \\
GS4-4583 & 1.71 $\pm$ 0.07 & 0.7 $\pm$ 0.1 & 0.67 $\pm$ 0.18 & 101 $\pm$ 21 & 1.77 $\pm$ 0.08 & 0.7 $\pm$ 0.1 \\
GS4-5217 & 1.50 $\pm$ 0.15 & 2.8 $\pm$ 0.5 & 0.58 $\pm$ 0.19 & 48 $\pm$ 17 & 1.69 $\pm$ 0.19 & 3.2 $\pm$ 0.5 \\
GS4-11016 & 1.86 $\pm$ 0.09 & 1.6 $\pm$ 0.2 & 0.38 $\pm$ 0.10 & 56 $\pm$ 7 & 1.96 $\pm$ 0.11 & 1.7 $\pm$ 0.2 \\
GS4-14747 & 2.24 $\pm$ 0.08 & 0.6 $\pm$ 0.1 & 0.85 $\pm$ 0.17 & 152 $\pm$ 32 & 2.31 $\pm$ 0.08 & 0.6 $\pm$ 0.1 \\
GS4-19348 & 4.07 $\pm$ 0.08 & 0.7 $\pm$ 0.0 & 0.85 $\pm$ 0.06 & 81 $\pm$ 12 & 4.22 $\pm$ 0.08 & 0.7 $\pm$ 0.0 \\
GS4-22825 & 1.34 $\pm$ 0.09 & 1.1 $\pm$ 0.2 & 0.65 $\pm$ 0.16 & 129 $\pm$ 16 & 1.43 $\pm$ 0.10 & 1.2 $\pm$ 0.2 \\
GS4-30274 & 1.62 $\pm$ 0.10 & 1.0 $\pm$ 0.3 & 0.72 $\pm$ 0.40 & 106 $\pm$ 46 & 1.64 $\pm$ 0.11 & 0.9 $\pm$ 0.2 \\
GS4-40185 & 0.36 $\pm$ 0.06 & $\cdots$ & $\cdots$ & $\cdots$ & $\cdots$ & $\cdots$ \\
GS4-41181 & 1.48 $\pm$ 0.07 & 0.5 $\pm$ 0.1 & 0.60 $\pm$ 0.41 & 151 $\pm$ 33 & 1.49 $\pm$ 0.08 & 0.5 $\pm$ 0.1 \\
GS4-44065 & 0.82 $\pm$ 0.11 & 2.2 $\pm$ 0.7 & 0.39 $\pm$ 0.23 & 103 $\pm$ 15 & 0.86 $\pm$ 0.13 & 2.3 $\pm$ 0.7 \\
GS4-45068 & 1.66 $\pm$ 0.08 & 0.7 $\pm$ 0.1 & 0.80 $\pm$ 0.14 & 64 $\pm$ 23 & 1.71 $\pm$ 0.08 & 0.7 $\pm$ 0.1 \\
GS4-45475 & 1.06 $\pm$ 0.24 & 5.0 $\pm$ 1.0 & $\cdots$ & $\cdots$ & 1.13 $\pm$ 0.31 & 5.1 $\pm$ 1.4
  \enddata
\end{deluxetable*}


\end{document}